\documentclass[12pt]{article}
\usepackage{graphicx,verbatim,array,multicol,palatino,amssymb,amsfonts, amsmath, epsfig,multirow,caption}
\usepackage{color}
\usepackage{psfrag}
\usepackage{chicago}
\usepackage{epstopdf}
\usepackage{pdflscape}
\usepackage{tikz}
\usepackage{subcaption}
\usepackage[subrefformat=parens,labelformat=parens]{subfig}
\usepackage{url}

\def\lboxit#1{\vbox{\hrule\hbox{\vrule\kern6pt
      \vbox{\kern6pt#1\kern6pt}\kern6pt\vrule}\hrule}}

\def\thick#1{\hbox{\rlap{$#1$}\kern0.25pt\rlap{$#1$}\kern0.25pt$#1$}}

\begin{document}
\title{A novel approach for Markov Random Field with intractable normalising constant on large lattices}
\author{W. Zhu \footnote{School of Mathematics and Statistics, University of New South Wales, Sydney 2052 Australia. Communicating Author Wanchuang Zhu: Email \tt{z3406438@unsw.edu.au.}} \: and
Y. Fan\footnote{School of Mathematics and Statistics, University of New South Wales, Sydney 2052 Australia.  Email \tt{Y.Fan@unsw.edu.au.}}
}

\maketitle
\begin{abstract}
The pseudo likelihood method of \shortciteN{besag1974spatial}, has remained a popular
method for estimating Markov random field on a very large lattice, despite various documented deficiencies. This is partly because it remains the only computationally tractable method for large lattices.  We introduce a novel method to estimate Markov random fields defined on a regular lattice.  The method takes advantage of conditional independence structures  and recursively decomposes a large lattice into smaller sublattices. An approximation is made at each decomposition. Doing so completely avoids the need to compute the troublesome normalising constant. The computational complexity is $O(N)$, where $N$ is the the number of pixels in lattice, making it computationally attractive for very large lattices. We show through simulation, that the proposed method performs well, even when compared to the methods using exact likelihoods.
\end{abstract}

\section{Introduction} \label{sec:intro}
Markov random field (MRF) models have an important role in modelling spatially correlated datasets. They have been used extensively in image and texture analyses ( \shortciteNP{nott1999pairwise}, \shortciteNP{hurn2003tutorial}),  image segmentation (\shortciteNP{pal1993review},  \shortciteNP{van1999automated},  \shortciteNP{celeux2003procedures}, \shortciteNP{li2009markov}), disease mapping (\shortciteNP{knorr2002block}, \shortciteNP{green2002hidden}), geostatistics (\shortciteNP{cressie1993statistics}) and more recently in social networks (\shortciteNP{everitt2012bayesian}).
In hidden Markov random field (HMRF) models, latent variables ${\bf z} = (z_1, \ldots, z_n)$ are introduced for each observed data $y_i, i=1,\ldots, n$,  where each pair $(y_i, z_i)$  has a corresponding spatial location. The MRF, and hence spatial interaction is modelled via ${\bf z}$ using an appropriate model, such as, Potts or autologistic models.

In what follows, we describe our proposed methodology in terms of the $q$-state Potts model,  although the method applies to other similar models, such as autologistic model and of course Ising model (a special case of Potts model when $q=2$). In the Bayesian framework, the distribution $\pi({\bf z} |\beta)$  can be seen as a prior distribution, and the hidden or missing observations $z_i, i=1,\ldots,n$ are treated as unknown parameters to be estimated. For instance, a common form of the posterior distribution of a $q$-component spatial mixture model takes the form
\begin{equation}
\pi({\bf z}, \beta,\theta|{\bf y}) \propto \prod_{i=1}^n \pi(y_i|\theta, z_i) \pi({\bf z}|\beta)\pi(\beta)\pi(\theta),
\end{equation}\label{eq:mix}
where $\pi(y_i|\theta, z_i)$ denotes the component distribution for $y_i$ conditional on the model parameters $\theta$ and $z_i$, $\pi(\theta)$ and $\pi(\beta)$ denote the prior and hyper prior for the unknown parameters.  Using the Potts model to define $ \pi({\bf z}|\beta)$, we have
\begin{equation}\label{eq:potts}
 \pi({\bf z}|\beta)=\frac{1}{\mathcal{C}({\beta})}\exp \{\beta \sum_{i \sim j} I(z_i=z_j) \},
\end{equation}
where $i\sim j$ indicates that $i$ and $j$ are neighbours, and $\mathcal{C}({\beta}) = \sum_{\bf z} \exp \{\beta \sum_{i \sim j} I(z_i=z_j) \}$ is the normalizing constant. $I(\cdot)$ is the indicator function, $I(z_i = z_j)=1$ if $z_i =z_j$ is true, otherwise $I(z_i = z_j)=0$. Figure \ref{simple mrf} (left panel) gives a pictorial illustration of a MRF with a first order neighbourhood structure, where each black site depends only on the four neighbouring gray sites on a 2D lattice. The 3D MRF is similarly defined with each site dependent on its neighbours on the left, right, front, back, above and below.  The parameter $\beta$ controls the degree of spatial dependence. See \shortciteN{wu1982potts} for more illustrations on the Potts model.

For relatively small random fields (less than $10 \times 10$), the normalizing constant $\mathcal{C}({\beta}) $  can be computed by summing exhaustively over all possible combinations of ${\bf z}$ for any given value of $\beta$. However, the calculation of  $\mathcal{C}({\beta})$ becomes computationally intractable for large spatial fields. The posterior distribution $\pi(\theta,{\bf z},\beta|y)$ is sometimes also referred to doubly-intractable distribution (\shortciteNP{murray06}).
This problem is well known in the statistical community, and has received considerable amount of attention in the literature, see \shortciteN{lynegass15} for a recent review.

\shortciteN{gelman1998simulating} used path sampling to directly approximate ratio of the normalizing constants, which can be used within posterior simulation algorithms such as MCMC, where only ratios are needed. Thermodynamic integration (TDI) is another approach which relies on Monte Carlo simulations.  \shortciteN{green2002hidden} for example adopted this approach by computing a look-up table offline.  Other simulation-based methods can be found in \shortciteN{geyer1992constrained}, \shortciteN{gu2001maximum}, \shortciteN{liang2007continuous} and references therein. However, most methods utilising Monte Carlo become computationally expensive for very large lattices.

The pseudo likelihood (PL) method of \shortciteN{besag1974spatial} approximates $\pi({\bf z}|\beta)$ as product of full conditional probabilities, where each term in the product is a full conditional of the neighbouring sites.
The normalizing constant for each term in the product then becomes trivial to compute. Note however, that this is a type of composite likelihood (\shortciteNP{lindsay1988composite}, \shortciteNP{varin2011overview}). The simplicity of the approach, coupled with its computational efficiency, makes the method still one of the most popular approaches in practice, particularly for large lattices. It has been noted in the literature that when the dependence is weak, the maximum pseudo-likelihood (MPLE) estimator behaves well and is almost efficient. In high dependence cases, the PL estimate is called into question, it has been shown to severely overestimate the dependence parameter, see \shortciteN{geyer1992constrained}.
\shortciteN{hurn2003tutorial} comments that that PL should only be considered for dependences below the critical value, and its effects on modelling data with long range dependences are not clear.
\shortciteN{cressie1998image} proposed a similar method known as partially ordered Markov models (POMMs), where the likelihood can be expressed as a product of conditional probabilities, without the need to compute the normalizing constant.
POMM defines parent sites for each point on the lattice, and the point only depends on its parents. However, only a subset of MRFs are expressible as POMMs.

 \shortciteN{reeves2004efficient} proposed a method for general factorizable models, which includes the autologistic and Potts model. This simple, yet effective approach is based on an algebraic simplification of the Markovian dependence structure, and is applicable to lattices with a small number of rows (up to 20).  As a result of the factorisation, the normalizing constant can be computed over the much smaller subsets of ${\bf z}$, making such computations feasible.
 \shortciteN{friel2009bayesian} extended the work of  \shortciteN{reeves2004efficient} to larger lattices by relaxing some of the dependence assumptions about
 $\pi({\bf z}|\beta)$, so that the full model is a product of factors, each of which is defined on sublattices computed using the method of \shortciteN{reeves2004efficient}. The sublattices are assumed to be independent, they term this reduced dependence approximation (RDA).
The authors showed that RDA can be efficiently applied to the binary MRF, but concluded that the extension to the Potts model may not be computationally tractable. Another similar idea can be found in \shortciteN{bartolucci2002recursive}, who also
presented a recursive algorithm using the product of conditional probabilities, their method is only applicable to lattices of up to 12 rows and columns.

Finally, another class of methods completely avoid the computation of the normalizing constant by ingeniously employing an auxiliary variable, see
\shortciteN{moller2006efficient}, \shortciteN{murray07advancesin}, \shortciteN{murray06}. However, the method is computationally very expensive, as well as requiring perfect simulation (\shortciteNP{proppwilson98}). \shortciteN{liang2010double} proposed a double Metropolis-Hastings sampler, in which the auxiliary variable is drawn more efficiently. More recently,  \shortciteN{liangjsl15} extended the exchange algorithm of  \shortciteN{murray06} to overcome the issue of obtaining perfect samples, using an importance sampling procedure coupled with a Markov chain running in parallel.   \shortciteN{everitt2012bayesian} proposed a sequential Monte Carlo method to deal with the same issue.

In many applications of MRFs, the size of the random field can be extremely large,  the rows and columns of the lattices are often in the order of hundreds or even thousands. In this article, we propose a new approach which is able to handle arbitrarily large lattices. Our approach takes advantage of the conditional independence structure of the MRF defined on a regular lattice, and recursively divides the field into smaller sub-MRFs. Each sub-MRF is then approximated by another Potts model, with weaker spatial interaction as the size of the grid on the lattice increases.

This paper is arranged as below.  Section \ref{sec: ICDA} presents our proposed method to tackle the intractable normalizing constant issue. Section~\ref{sec:Generalization} describes
the generalisation of the proposed methodology to second (or higher) order neighbourhood structures. Section \ref{sec:validation} provides extensive simulation studies of our proposed methodology, and makes some comparisons with existing methodology.  A real data application is presented in Section \ref{sec:application}. Finally, Section \ref{sec:conclusion} concludes with some discussions.

\section{A recursive decomposition method}
\label{sec: ICDA}
\begin{figure} [ht]
\centering
\begin{tikzpicture}
 \draw [step=0.5cm,gray,very thin] (0,0) grid (3.5,3.5);
 \foreach \x in {0,1,2,3}
 \foreach \y in {0,1,2,3}
 { \filldraw [black] (\x,\y) circle  (2pt) ;
 }

 \foreach \x in {0.5,1.5,2.5,3.5}
 \foreach \y in {0.5,1.5,2.5,3.5}
 { \filldraw [black] (\x,\y) circle  (2pt) ;
 }

 \foreach \x in {0.5,1.5,2.5,3.5}
 \foreach \y in {0,1,2,3}
 { \filldraw [gray] (\x,\y) circle  (2pt) ;
 }

  \foreach \x in {0,1,2,3}
 \foreach \y in  {0.5,1.5,2.5,3.5}
 { \filldraw [gray] (\x,\y) circle  (2pt) ;
 }

 \begin{scope} [xshift=5cm]
 \draw [step=0.5cm,very thin,gray] (0,0) grid (3.5,3.5);
 \foreach \x in {0,1,2,3}
 \foreach \y in {0,1,2,3}
 { \filldraw [black] (\x,\y) circle  (2pt) ;
 }

 \foreach \x in {0.5,1.5,2.5,3.5}
 \foreach \y in {0.5,1.5,2.5,3.5}
 { \filldraw [black] (\x,\y) circle  (2pt) ;
 }
 \end{scope}

 \begin{scope} [xshift=10cm]
 \draw [step=0.5cm,very thin,gray] (0,0) grid (3.5,3.5);
 \foreach \x in {0,1,2,3}
 \foreach \y in {0,1,2,3}
 { \filldraw [black] (\x,\y) circle  (2pt) ;
 }

 \foreach \x in {0.5,1.5,2.5,3.5}
 \foreach \y in {0.5,1.5,2.5,3.5}
 { \filldraw [gray] (\x,\y) circle  (2pt) ;
 }
 \end{scope}

\end{tikzpicture}
\caption{Left panel: a first order neighbourhood MRF, with black and grey points depicting ${\bf z}$.  Each site only depends on the nearest four neighbours of the other color. Middle panel:  the sub lattice ${\bf z}^{(2)}$. Right panel: ${\bf z}^{(2)}$  further divided into two parts based on the first order neighbourhood.}
\label{simple mrf}
\end{figure}
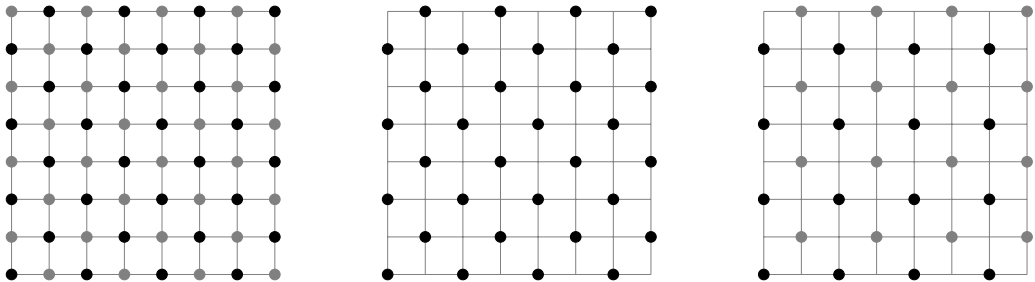

Consider the first order neighbourhood structure defining the MRF.  The left panel of Figure \ref{simple mrf} depicts the location of the
latent variable ${\bf z}$ defined on a regular lattice with a first order neighbourhood dependence structure. Here each black site depends only on its neighbouring grey sites. A natural consequence of this dependence structure is that, given the black sites, all the grey sites are independent, and vice versa.  Thus conditioning on the grey sites, and decomposing the Potts model of Equation \eqref{eq:potts} we have
\begin{equation}\label{eqn:decomp}
\pi({\bf z}|\beta)\equiv \pi_{potts}({\bf z}|\beta) = \pi({\bf z}^{(1)}|{\bf z}^{(2)}, \beta)\pi({\bf z}^{(2)}|\beta),
\end{equation}
where ${\bf z}^{(1)}$ corresponds to the grey sites in Figure \ref{simple mrf}, left panel. The conditional independence property allows us to compute $\pi({\bf z}^{(1)}|{\bf z}^{(2)}, \beta)$ directly as
\begin{equation}\label{eqn:indeplik}
\pi({\bf z}^{(1)}|{\bf z}^{(2)}, \beta) =\prod_{i=1}^{n_1} \frac{\exp \{\beta \sum_{ i \sim j} I(z^{(1)}_i = z^{(2)}_j) \}}{\sum_{z_i^{(1)}=1,\ldots,q} \exp \{\beta \sum_{ i\sim j} I(z^{(1)}_i = z^{(2)}_j)\}}
\end{equation}
producting over all $n_1$ observations in ${\bf z}^{(1)}$.\\

The field ${\bf z}^{(2)}$ is depicted by the middle panel in Figure \ref{simple mrf}. Here we approximate the dependence structure of this sub-MRF with another MRF model using the first order neighbourhood, as seen in the right panel of Figure \ref{simple mrf}. The dependence in ${\bf z}^{(2)}$ is weaker than the original MRF as the sites are further away from each other. Thus we approximate ${\bf z}^{(2)}$  again as a Potts model with first order neighbourhood $\pi_{potts}({\bf z}^{(2)} |\alpha\beta)$. That is,
\begin{equation}\label{eqn:pottsdecay}
\pi({\bf z}^{(2)}|\beta) \approx \pi_{potts}({\bf z}^{(2)} |\alpha\beta)
\end{equation}
with decay coefficient $0\leq \alpha \leq 1$. Related references on long-range decay  in spatial interactions can be found in \shortciteN{kosterlitz1974critical}, \shortciteN{wu1982potts}, \shortciteN{aizenman1988discontinuity} and \shortciteN{luijten1995monte}.

If the field in ${\bf z}^{(2)}$ is large, then we can apply the same principle to ${\bf z}^{(2)}$, as in Equation \eqref{eqn:decomp}, to obtain  ${\bf z}^{(3)}$ and ${\bf z}^{(4)}$, and so on. Until we end up with a Potts field for which computation for its normalizing constant becomes trivial. Hence,  our approximation to the original Potts model by splitting the MRF into 2$T$ fields is given by
\begin{equation}\label{eqn:RcoDA}
\pi_{potts}({\bf z}|\beta)\approx  \left\{ {\prod_{i \in  \mathcal{I}}} \pi({\bf z}^{(i)}|{\bf z}^{(i+1)}, \alpha^{(i-1)/2}\beta)\right\}\pi_{potts}({\bf z}^{(2T)}| \alpha^{T}\beta),
\end{equation}
where ${\mathcal{I} =\{ 1,3,\cdots, 2T-1\}}$. When $T=0$, Equation (\ref{eqn:RcoDA}) degenerates to the original Potts model. We term this approximation as recursive conditional decomposition approximation (RCoDA).
In the approximation above only the last term needs the calculation of the normalizing constant, which is easy for small fields.

Computational tractability dictates that we choose value of splits $T$, such that  the Potts term on the right hand side of Equation \eqref{eqn:RcoDA} becomes small enough to be tractable. Simulation studies for varying $T$ over a range of values of $\beta$ showed that the results are largely insensitive to the choice of $T$. In practice, we can choose $T$ so that the size of ${\bf z}^{(2T)}$ is no larger than 4 $\times$ 4. Note also that in relatively large fields with  weaker spatial dependences, resulting in a large number of $T$, the factor $\alpha^{T-1}$ tends to zero. In these cases, the term $\pi_{potts}({\bf z}^{(2T)}| \alpha^{T-1}\beta)$ in Equation \eqref{eqn:RcoDA}  can be treated as an independent random field.

Equation \eqref{eqn:RcoDA} can be viewed as an approximation to the $q$-state Potts model. Alternatively, one can also view this model as being more flexible than the standard Potts model, particularly when one is interested in understanding different types of decay in the dependence when long range dependence is present. It is possible to model the rate of decay differently to what is considered in this paper.
Here the dependence at the $T^{th}$ sublattice is modelled as $\alpha^T\beta$, where $\beta$ is the global dependence parameter. This rate of decay was found by several authors in several difference applications, see \shortciteN{kosterlitz1974critical}, \shortciteN{wu1982potts}. We will investigate this assumption more closely in Section \ref{sec:validation}.
Another important question when an approximation is used in place of the true likelihood, is whether this yields valid inference. \shortciteN{monahanboos92} introduces the notion of validity of posterior inference based on the correct coverage probability. We will also validate the use of RCoDA under this notion in
Section \ref{sec:validation}.

\section{Extensions to second order structure}
\label{sec:Generalization}

\begin{figure} [ht]
 \centering
 \begin{minipage}[b]{.3\linewidth}
    \centering
   \begin{tikzpicture} [scale=1.2]
  \draw [step=0.5cm][gray,very thin] (0,0) grid (2,2);
 \foreach \x in {2/2}
 \foreach \y in {2/2}
 { \filldraw [black] (\x,\y) circle  (1.5pt) ;
 }
 \foreach \x in {1/2,3/2}
 \foreach \y in {1/2,3/2}
 { \filldraw [gray] (\x,\y) circle  (1.5pt) ;
 }

 \foreach \x in {0.5,1.5}
 \foreach \y in {1}
 { \filldraw [gray] (\x,\y) circle  (1.5pt) ;
 }

  \foreach \x in {1}
 \foreach \y in  {0.5,1.5}
 { \filldraw [gray] (\x,\y) circle  (1.5pt)  ;
 }

 \end{tikzpicture}
    \subcaption{Second order structure in 2D MRF. Gray sites are neighbourhoods of the black site. \\ \\}
  \end{minipage}
  \hfill
\begin{minipage}[b]{.3\linewidth}
  \centering
  \begin{tikzpicture}

\foreach \x in {1.5,2.5}
\foreach \y in {1.5,2.5}
\foreach \z in {1,2,3}
        \draw [gray,very thin](\x,\y,\z) +(-.5,-.5) rectangle ++(.5,.5);
\foreach \x in {1,2,3}
        \draw [gray,very thin](\x,1,1) -- (\x,1,3) (\x,2,1) -- (\x,2,3) (\x,3,1) -- (\x,3,3)  ;

\foreach \x in {1,2,3}
\foreach \y in {1,2,3}
\foreach \z in {1,2,3}
        \filldraw [gray](\x,\y,\z) circle (1.5pt);
\foreach \x in {1,3}
\foreach \y in {1,3}
\foreach \z in {1,3}
         \filldraw [black] (\x,\y,\z) circle (1.5pt);

\filldraw [black] (2,2,2) circle (1.5pt);

 \end{tikzpicture}
 \subcaption{The first type of second order structure in 3D lattice: 18 neighbourhoods structure. All the gray sites are neighbourhoods of the black sites.}
\end{minipage}
\hfill
\begin{minipage}[b]{.3\linewidth}
  \centering
  \begin{tikzpicture}

\foreach \x in {1.5,2.5}
\foreach \y in {1.5,2.5}
\foreach \z in {1,2,3}
        \draw [gray,very thin](\x,\y,\z) +(-.5,-.5) rectangle ++(.5,.5);
\foreach \x in {1,2,3}
        \draw [gray,very thin](\x,1,1) -- (\x,1,3) (\x,2,1) -- (\x,2,3) (\x,3,1) -- (\x,3,3)  ;

\foreach \x in {1,2,3}
\foreach \y in {1,2,3}
\foreach \z in {1,2,3}
        \filldraw [gray](\x,\y,\z) circle (1.5pt);

\filldraw [black] (2,2,2) circle (1.5pt);
%
 \end{tikzpicture}
 \subcaption{Second type of second order structure in 3D lattice: 26 neighbourhoods structure. All the gray sites are neighbouhoods of the black site.}
\end{minipage}
 \caption{Second order structure in multiple scenarios.}
 \label{secondorder}
\end{figure}
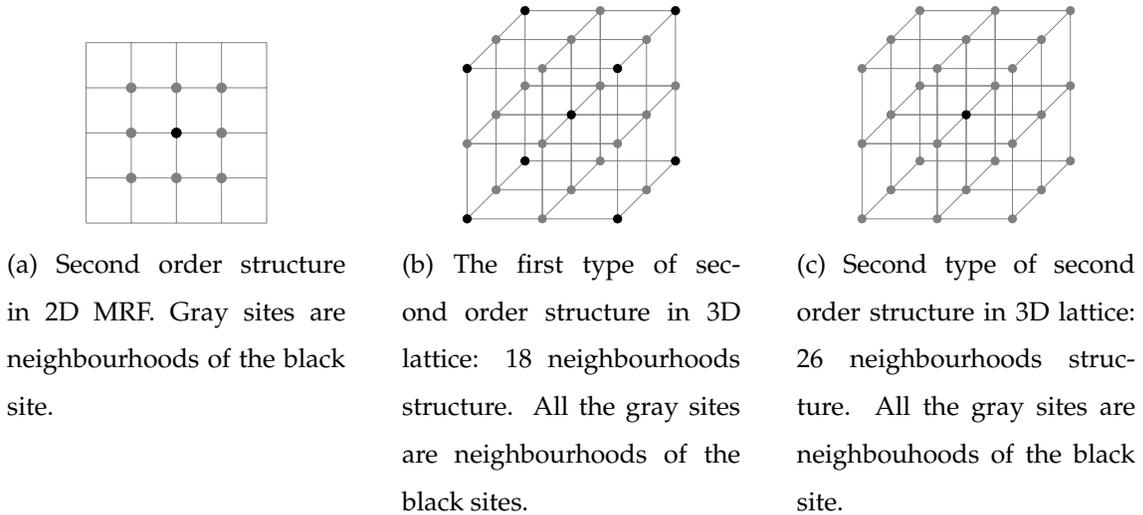

The most common neighbourhood structures in MRFs are the first and second order \shortcite{besag1974spatial}.
 One of the most common types of second order structure for 2D MRFs is shown in Figure \ref{secondorder}(a), where each site has six neighbours. There are multiple types of second order structures for 3D MRFs. Figure \ref{secondorder}(b) and Figure \ref{secondorder}(c) show the 18 and 26 neighbourhood structures in 3D.
Our proposed methodology requires that we split the entire lattices into non-overlapping sublattices. Here we use the "coding method" approach to obtain the sublattices (see \shortciteN{besag1974spatial},  \shortciteN{winkler2003image} and \shortciteN{wilkinson2006parallel}). The minimum number of sublattices for  a first order structures is 2 in both 2D and 3D lattices, and 4, 4 and 8 in second order neighbourhoods structures with 8 neighbours in 2D, 18 neighbours in 3D and 26 neighbours in 3D respectively. These numbers are the so-called chromatic number, more details on these can be found in \shortciteN{feng2008bayesian} and \shortciteN{feng2012mri}.

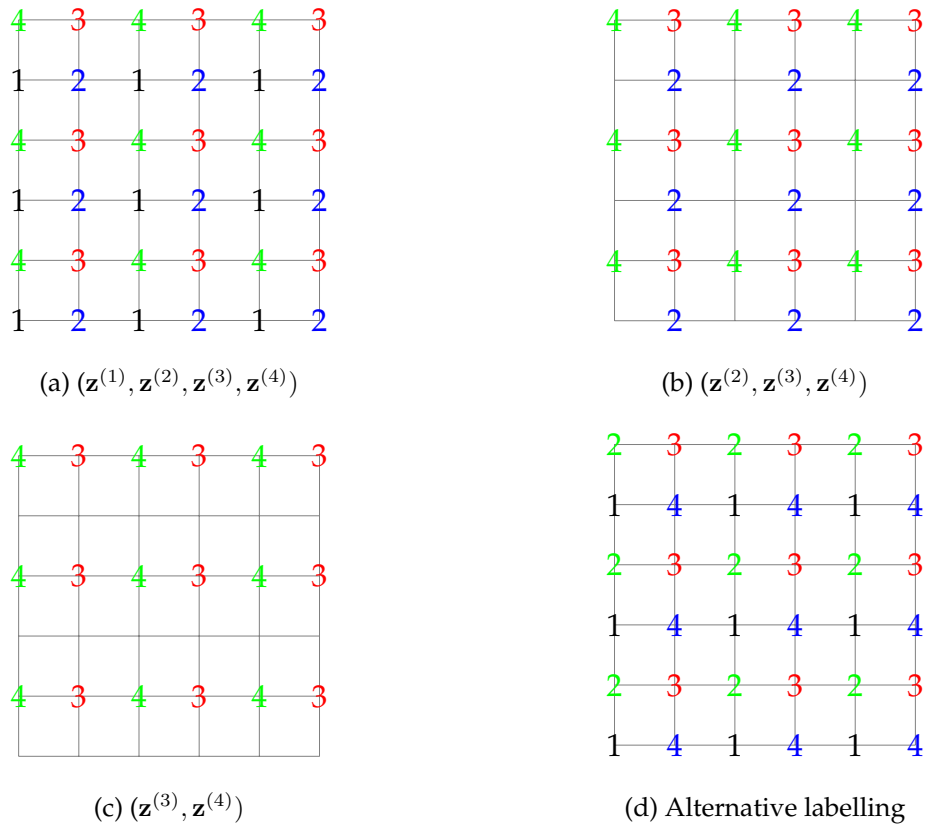
\begin{figure} [ht]

\begin{tabular}{cc}
\begin{minipage}[b]{.5\linewidth}
 \centering
 \begin{tikzpicture} [scale=0.8]
  \draw [step=1cm][gray,very thin] (0,0) grid (5,5);
 \foreach \x in {0,2,4}
 \foreach \y in {0,2,4}
 { 
 \draw  (\x,\y) node   {\textcolor {black}{1}};
 }

 \foreach \x in {1,3,5}
 \foreach \y in {0,2,4}
          {
          \draw  (\x,\y) node   {\textcolor {blue}{2}};}

 \foreach \x in {1,3,5}
 \foreach \y in {1,3,5}
 { 
 \draw  (\x,\y) node   {\textcolor {red}{3}};

 }

 \foreach \x in {0 ,2,4}
 \foreach \y in {1,3,5}
 { 
 \draw  (\x,\y) node   {\textcolor {green}{4}};
 }

 \end{tikzpicture}
 \subcaption{(${\bf z}^{(1)}, {\bf z}^{(2)}, {\bf z}^{(3)}, {\bf z}^{(4)})$}
 \label{fig:4parts}
  \end{minipage} &

\begin{minipage}[b]{.5\linewidth}
 \centering
 \begin{tikzpicture} [scale=0.8]
  \draw [step=1cm][gray,very thin] (0,0) grid (5,5);
 \foreach \x in {0,2,4}
 \foreach \y in {0,2,4}
 { 
 \draw  (\x,\y) node   {\textcolor {black}{}};
 }

 \foreach \x in {1,3,5}
 \foreach \y in {0,2,4}
          {
          \draw  (\x,\y) node   {\textcolor {blue}{2}};}

 \foreach \x in {1,3,5}
 \foreach \y in {1,3,5}
 { 
 \draw  (\x,\y) node   {\textcolor {red}{3}};

 }

 \foreach \x in {0 ,2,4}
 \foreach \y in {1,3,5}
 { 
 \draw  (\x,\y) node   {\textcolor {green}{4}};
 }

 \end{tikzpicture}
  \subcaption{(${\bf z}^{(2)}, {\bf z}^{(3)}, {\bf z}^{(4)})$}
 \label{fig:3parts}
  \end{minipage} \\

\begin{minipage}[b]{.5\linewidth}
\centering
 \begin{tikzpicture} [scale=0.8]

  \draw [step=1cm][gray,very thin] (0,0) grid (5,5);
 \foreach \x in {0,2,4}
 \foreach \y in {0,2,4}
 { 
 \draw  (\x,\y) node   {\textcolor {black}{}};
 }

 \foreach \x in {1,3,5}
 \foreach \y in {0,2,4}
          {
          \draw  (\x,\y) node   {\textcolor {blue}{}};}

 \foreach \x in {1,3,5}
 \foreach \y in {1,3,5}
 { 
 \draw  (\x,\y) node   {\textcolor {red}{3}};

 }

 \foreach \x in {0 ,2,4}
 \foreach \y in {1,3,5}
 { 
 \draw  (\x,\y) node   {\textcolor {green}{4}};
 }
 \end{tikzpicture}
 \subcaption{(${\bf z}^{(3)}, {\bf z}^{(4)})$}
 \label{fig:2parts}
 \end{minipage} &

 \begin{minipage}[b]{.5\linewidth}
 \centering
 \begin{tikzpicture} [scale=0.8]
  \draw [step=1cm][gray,very thin] (0,0) grid (5,5);
 \foreach \x in {0,2,4}
 \foreach \y in {0,2,4}
 { 
 \draw  (\x,\y) node   {\textcolor {black}{1}};
 }

 \foreach \x in {1,3,5}
 \foreach \y in {0,2,4}
          {
          \draw  (\x,\y) node   {\textcolor {blue}{4}};}

 \foreach \x in {1,3,5}
 \foreach \y in {1,3,5}
 { 
 \draw  (\x,\y) node   {\textcolor {red}{3}};

 }

 \foreach \x in {0 ,2,4}
 \foreach \y in {1,3,5}
 { 
 \draw  (\x,\y) node   {\textcolor {green}{2}};
 }

 \end{tikzpicture}
 \subcaption{Alternative labelling}
 \label{fig:collabel}
 \end{minipage}  \\
\end{tabular}
 \caption{(a) Using the coding method approach, a $6\times 6$ lattice is split into 4 sublattices. Each sublattice is labelled by corresponding number. (b) Sublattices with ${\bf z}^{(1)}$ removed. (c) Sublattice of ${\bf z}^{(3)}, {\bf z}^{(4)}$. (d) Alternative labelling, swapping 2 with 4 in (a). }
 \label{4sublattices}
\end{figure}

Focusing on the case of second order neighbourhood in 2D, we proceed by first identifying the 4 sublattices using the coding method. Figure \ref{4sublattices}(a) shows the corresponding lattice being split into 4 sublattices, corresponding to $({\bf z}^{(1)}, {\bf z}^{(2)}, {\bf z}^{(3)}, {\bf z}^{(4)})$. Following the same decomposition as in Equation \eqref{eqn:decomp}, we obtain
\begin{equation}\label{eqn:decomp2}
\pi({\bf z}|\beta)=\pi({\bf z}^{(1)}|{\bf z}^{(2)}, {\bf z}^{(3)}, {\bf z}^{(4)}, \beta)\pi({\bf z}^{(2)}|{\bf z}^{(3)}, {\bf z}^{(4)},\beta)\pi({\bf z}^{(3)}, {\bf z}^{(4)}|\beta).
\end{equation}
The first term on the right hand side of Equation \eqref{eqn:decomp2} can be estimated as product of full conditionals
similarly to Equation \eqref{eqn:indeplik}, see Figure \ref{4sublattices}(a) for the neighbourhood of ${\bf z}^{(1)}$.

The second term $\pi({\bf z}^{(2)}|{\bf z}^{(3)}, {\bf z}^{(4)},\beta)$ cannot be computed exactly, see Figure \ref{4sublattices}(b) for a pictorial depiction of the field for $({\bf z}^{(2)}, {\bf z}^{(3)}, {\bf z}^{(4)})$. This term is the marginal likelihood of the second order neighbourhood Potts model with ${\bf z}^{(1)}$ integrated out, and would be as difficult to compute as the original problem.
We consider two types of approximations for this term. In our first approximation, we assume conditional independence between ${\bf z}^{(1)}$ and ${\bf z}^{(2)}$,
thus allowing $\pi({\bf z}^{(2)}|{\bf z}^{(3)}, {\bf z}^{(4)},\beta)$ to be computed similarly to the first term, producting over all conditionally independent terms. We term this approach as RCoDA marginal (RCoDA-M).
In our second approximation, using a  similar approach to pseudo-likelihood approaches, we re-write the first two terms on the right hand side of Equation \eqref{eqn:decomp2} as
\begin{equation}\label{eqn:decomp2rw}
\pi({\bf z}^{(1)},{\bf z}^{(2)}| {\bf z}^{(3)}, {\bf z}^{(4)}, \beta)=\pi({\bf z}^{(1)}|{\bf z}^{(2)}, {\bf z}^{(3)}, {\bf z}^{(4)}, \beta)\pi({\bf z}^{(2)}| {\bf z}^{(1)},{\bf z}^{(3)}, {\bf z}^{(4)}, \beta),
\end{equation}
where both terms on the right hand side can be computed easily due to the conditional independence properties of these two subfields.  We term this approach as RCoDA conditional (RCoDA-C).

Finally, the remaining field involving only $({\bf z}^{(3)}, {\bf z}^{(4)})$ (as shown in Figure  \ref{4sublattices}(c)), can again be approximated by a second order neighbourhood Potts model of the form $\pi_{potts}({\bf z}^{(3)}, {\bf z}^{(4)}|\alpha\beta)$. This is done similarly to the first order case, and again modelling the spatial correlation with a decay term $\alpha$.  Note that the distances  between sites only increase either between rows, or columns depending on the iteration of the recursion.  To overcome this issue, we use alternate labelling  between each iteration of the recursion. For example, the labels between 2 and 4 are swapped in \ref{4sublattices}(d) after each recursion, increasing the distance between columns after this iteration. In summary, for every two iterations the distances change uniformly over the entire field.

Although more complicated neighbourhood structures work under the same principle, their conditional independence structures may not be as easy to take advantage of, especially those with higher chromatic numbers.

\section{Simulation study}\label{sec:validation}
In this section we perform extensive simulation studies to validate the proposed approach. Where possible, we compare our results with other existing methods.
Simulations are performed for both first and second order neighbourhoods defined on a regular 2D lattice.

\subsection{First order neighbourhood}
We first evaluate the performance of our estimation of $\beta$, for the first order neighbourhood dependences. We consider 2D lattices of sizes 32$\times$32, 128$\times$128 and 256$\times$256. It is well known that the Potts model exhibits the so called
phase transition, where for $\beta > \beta_{crit}$, the model will transit from disordered to ordered pattern or phase. This means that the sites will eventually all be in the same state as $\beta$ increases.  For a general $q$-state model, the precise value of the critical value is difficult to determine. For the Ising model ($q=2$) defined over 2D lattice,  \shortciteN{potts1952some} suggests setting $\beta_{crit}= \log(1+\sqrt{q})$, with $\beta_{crit}\approx 0.88$ for $q=2$ and $\beta_{crit}\approx 1.01$ for $q=3$.  \shortciteN{barkema1991numerical} suggests setting the critical values to 0.44 for $q=2$ and  0.503 for $q=3$. This is not compatible with the conclusion in \shortciteN{potts1952some} because they use different definitions of Potts model. In \shortciteN{barkema1991numerical} Potts model is defined as $\pi({\bf z}|\beta)= \exp\{ \beta \sum_{i \sim j} z_i z_j \}/ \mathcal{C}(\beta)$, where $z_i \in \{-1, 1\}$. This defination is different with Equation \ref{eq:potts}. But in essence, they have
same conclusion on critical value. Here we will restrict our analyses to $\beta$ below the critical values recommended by \shortciteN{potts1952some}, and consider the set of values $0.1, 0.2,\ldots, 0.8$ for $\beta$ .

For each value of $\beta$, we simulated  200 replicate datasets from the $q$-state Potts model using MCMC. Data from the Potts model was generated using Gibbs sampling using purpose written codes in Matlab. The final iterate after 5000 MCMC steps was then used as the observed data from the Potts model. Throughout our implementations of RCoDA, the priors $\beta \sim U(0, 0.9)$ and $\alpha \sim U(0,1)$ were used, and MCMC was used to obtain posterior estimates for both $\alpha$ and $\beta$. Approximately 6000 iterations with the first 2000 iterations as burn in were sufficient to obtain convergence for all models implemented.  For lattices of sizes 32$\times$32, 128$\times$128 and 256$\times$256, we decomposed the field until the smallest one is  4$\times$4, corresponding to $T=6,10,12$ respectively for the three different sized lattices. For comparison, we also implemented PL (\shortciteNP{besag1974spatial}),  TDI (\shortciteNP{green2002hidden}) and RDA (\shortciteNP{friel2009bayesian}) methods.  With the
exception of RDA, all methods were implemented in Matlab, RDA
was implemented using
the modified codes kindly provided by the authors. RDA was only implemented for the small field with $q=2$, as the method was developed for $q=2$, and the codes were also not available for larger lattice sizes.

\begin{table}[ht] \small
 \centering
   \begin{tabular}{|c|c|c|c|c|c|c|c|c|c|c|c|}
    \hline
    \hline
       $\beta$ &&& 0.100& 0.200&0.300& 0.400& 0.500& 0.600& 0.700&0.800\\
    \hline
    \multirow{4}{*}{$32^2$}& \multirow{2}{*}{RCoDA}  
                    &q=2 & 0.039 & 0.047 & 0.048 & 0.053 & 0.053 & 0.057 & 0.057 & 0.051\\
                & &q=3    & 0.039 & 0.047 & 0.051 & 0.051 & 0.053 & 0.049 & 0.049 & 0.046\\
    \cline{2-11}
%
 &   \multirow{2}{*} {PL}   
               &q=2   &  0.043 & 0.046 & 0.044 & 0.049 & 0.048 & 0.046 & 0.046 & 0.053\\
		& &q=3  &  0.044 & 0.046 & 0.047 & 0.049 & 0.051 & 0.044 & 0.042 & 0.047\\
     \cline{2-11}
 &   \multirow{2}{*} {TDI}  
             &q=2   &0.040 & 0.042 & 0.042 & 0.043 & 0.038 & 0.037 & 0.036 & 0.032  \\
             &&q=3 &0.040 & 0.044 & 0.045 & 0.045 & 0.045 & 0.039 & 0.034 & 0.034  \\
     \cline{2-11}
&    \multirow{2}{*} {RDA}    
             &q=2  & 0.040& 0.043& 0.042& 0.043& 0.038& 0.037& 0.036& 0.032\\
         &&q=3 & - & - & - & - & - &- &- &- \\
     \hline
      \multirow{2}{*}{$128^2$}&\multirow{2}{*}{RCoDA}
               &q=2  & 0.011 & 0.012 & 0.011 & 0.014 & 0.012 & 0.014 & 0.016 & 0.017 \\
                &&q=3 & 0.011 & 0.012 & 0.011 & 0.011 & 0.011 & 0.012 & 0.012 & 0.013\\
%
     \cline{2-11}

  &\multirow{2}{*}{PL}
              &q=2 & 0.011 & 0.011 & 0.011 & 0.012 & 0.011 & 0.011 & 0.012 & 0.012  \\
              &&q=3 &  0.011 & 0.012 & 0.011 & 0.011 & 0.011 & 0.010 & 0.011 & 0.011 \\
     \hline

     \cline{2-11}

  &\multirow{2}{*}{TDI}
              &q=2 &   0.011 & 0.011 & 0.010 & 0.010 & 0.009 & 0.009 & 0.008 & 0.007 \\
              &&q=3 &  0.011 & 0.011 & 0.010 & 0.010 & 0.010 & 0.009 & 0.008 & 0.008 \\
     \hline

  \multirow{2}{*}{$256^2$}&\multirow{2}{*}{RCoDA}
            &q=2 &0.006 & 0.005 & 0.006 & 0.006 & 0.006 & 0.008 & 0.009 & 0.013 \\
            &&q=3 & 0.006 & 0.006 & 0.005 & 0.007 & 0.006 & 0.006 & 0.007 & 0.007\\
           \cline{2-11}
  &\multirow{2}{*}{PL}
            &q=2 & 0.006 & 0.005 & 0.006 & 0.006 & 0.005 & 0.006 & 0.006 & 0.006\\
            &&q=3 &  0.006 & 0.005 & 0.006 & 0.006 & 0.005 & 0.006 & 0.006 & 0.006 \\
         \cline{2-11}

  &\multirow{2}{*}{TDI}
              &q=2 &  0.005 & 0.005 & 0.006 & 0.005 & 0.005 & 0.004 & 0.004 & 0.004 \\
              &&q=3 &  0.006 & 0.006 & 0.005 & 0.005 & 0.005 & 0.005 & 0.004 & 0.004 \\

     \hline \hline
   \end{tabular}
   \caption{Root mean squared error of $\beta$ for a first order neighbourhood dependence.  Based on 200 simulated data sets for each 32$\times$32, 128$\times$128 and 256$\times$256  lattices.  $q=2$ and $q=3$.}
 \label{validate}
\end{table}

Table \ref{validate} shows the root mean squared error of the $\beta$ estimation for  $q=2$ and $q=3$, for lattice sizes of $32 \times 32$, $128 \times 128$ and $256 \times 256$. The results are very similar for the different values of $q$ and $\beta$. While all the methods obtained small root mean squared error estimates, the performances in larger lattices between the different methods were almost indistinguishable.

We have also investigated the effects of using different values of $T$, i.e., the number of times to split the random field, and again the results were broadly insensitive to this specification. Numerical results are omitted from presentation here.

\begin{figure}[ht]
 \centering
  \includegraphics[width=3.5cm,height=6cm]{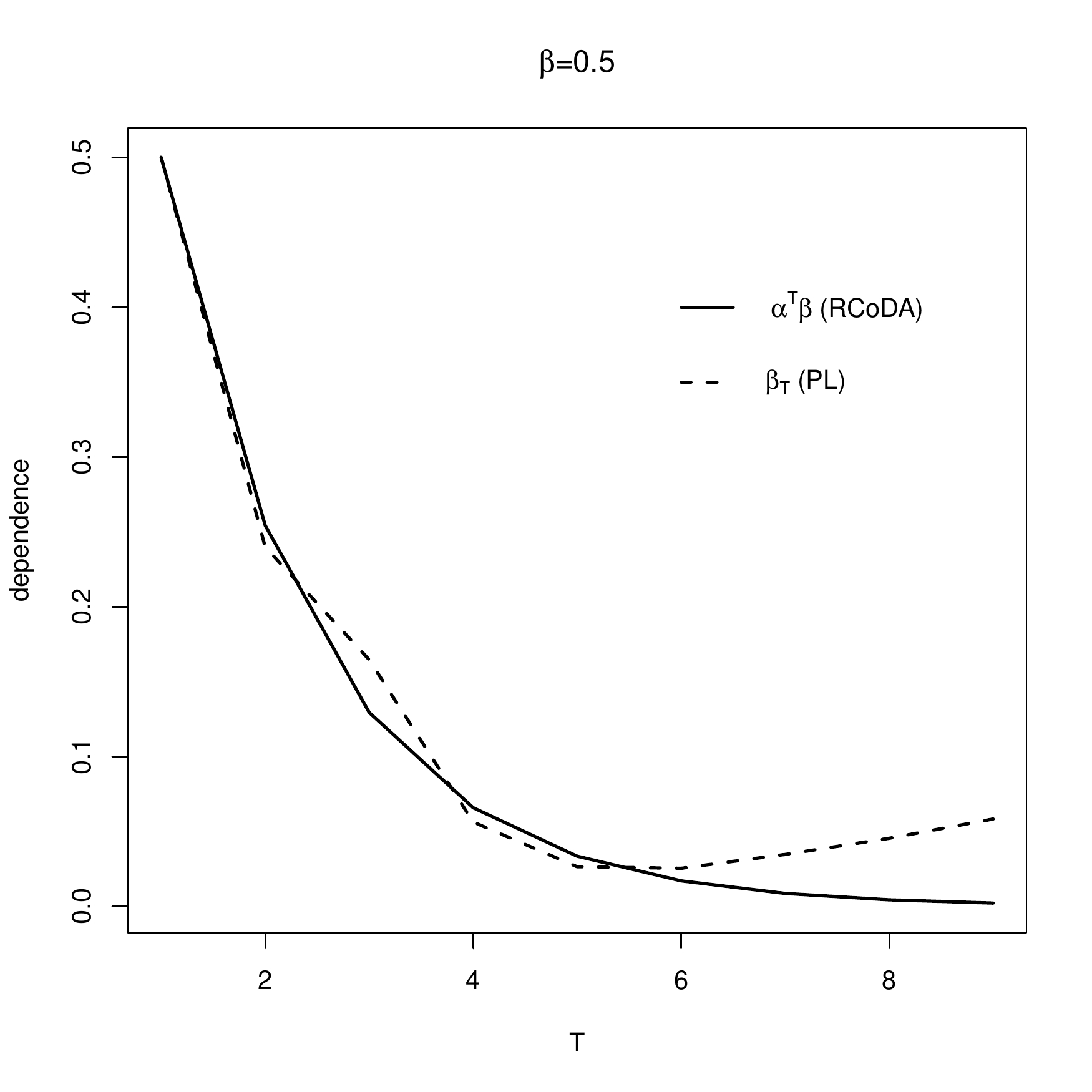}
    \includegraphics[width=3.5cm,height=6cm]{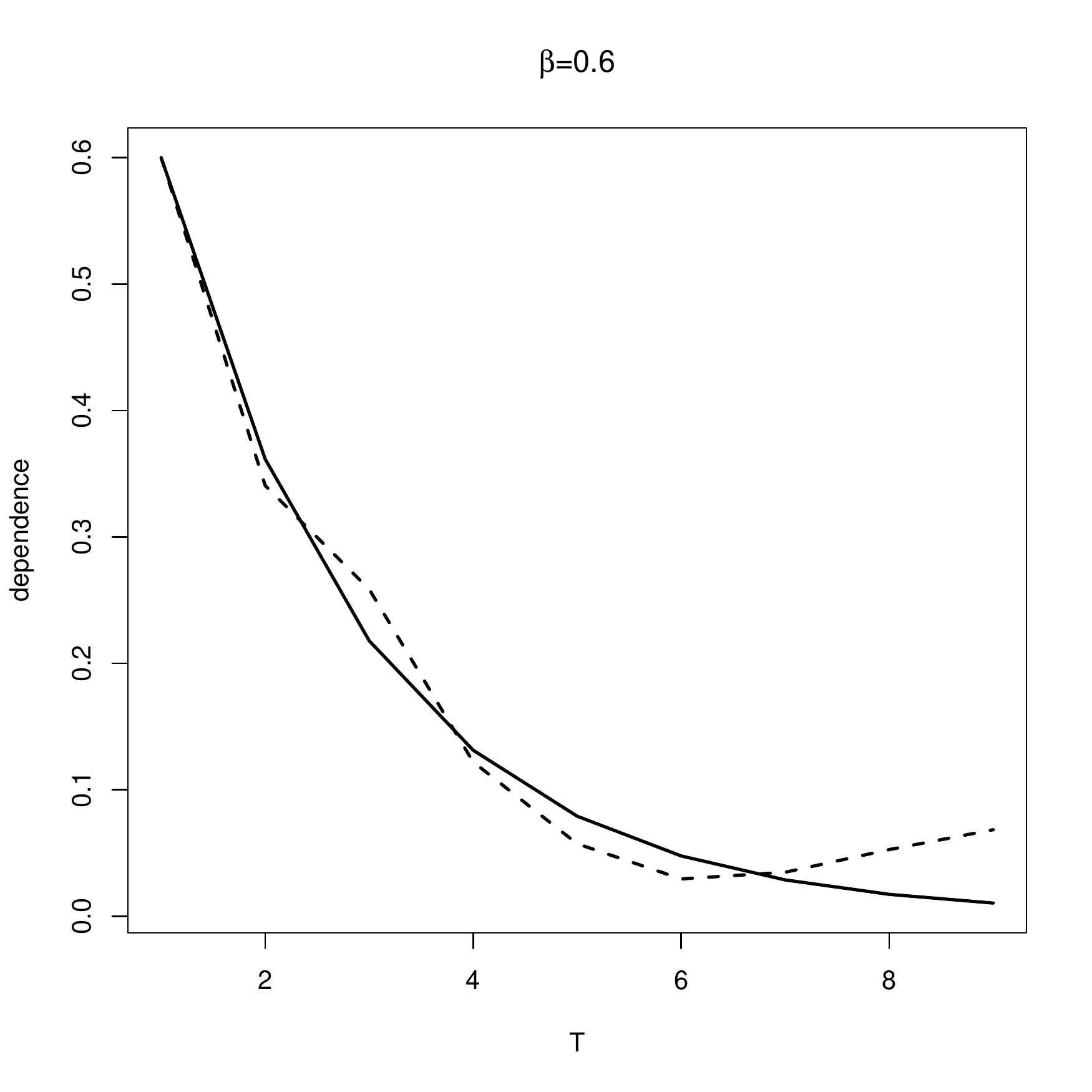}
    \includegraphics[width=3.5cm,height=6cm]{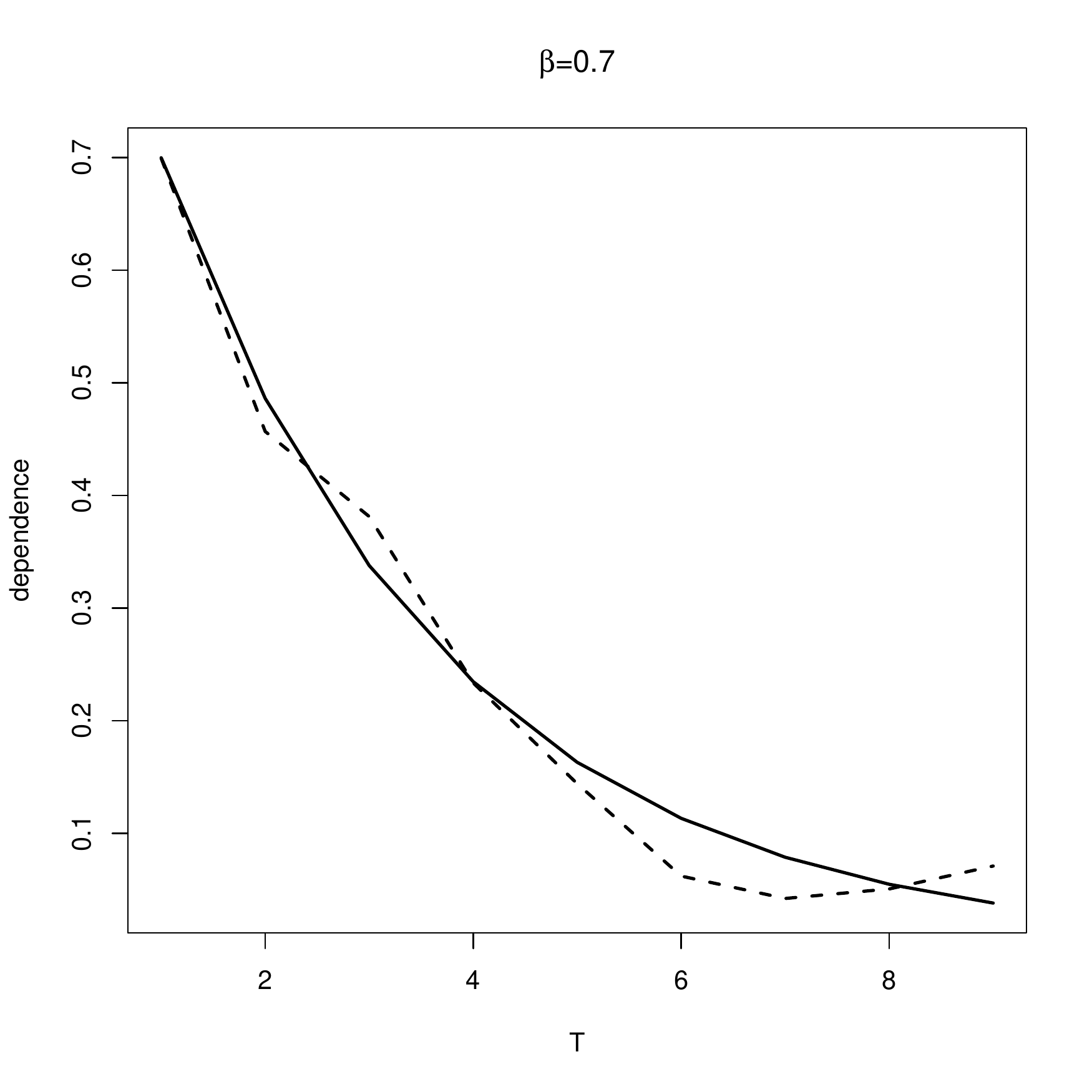}
    \includegraphics[width=3.5cm,height=6cm]{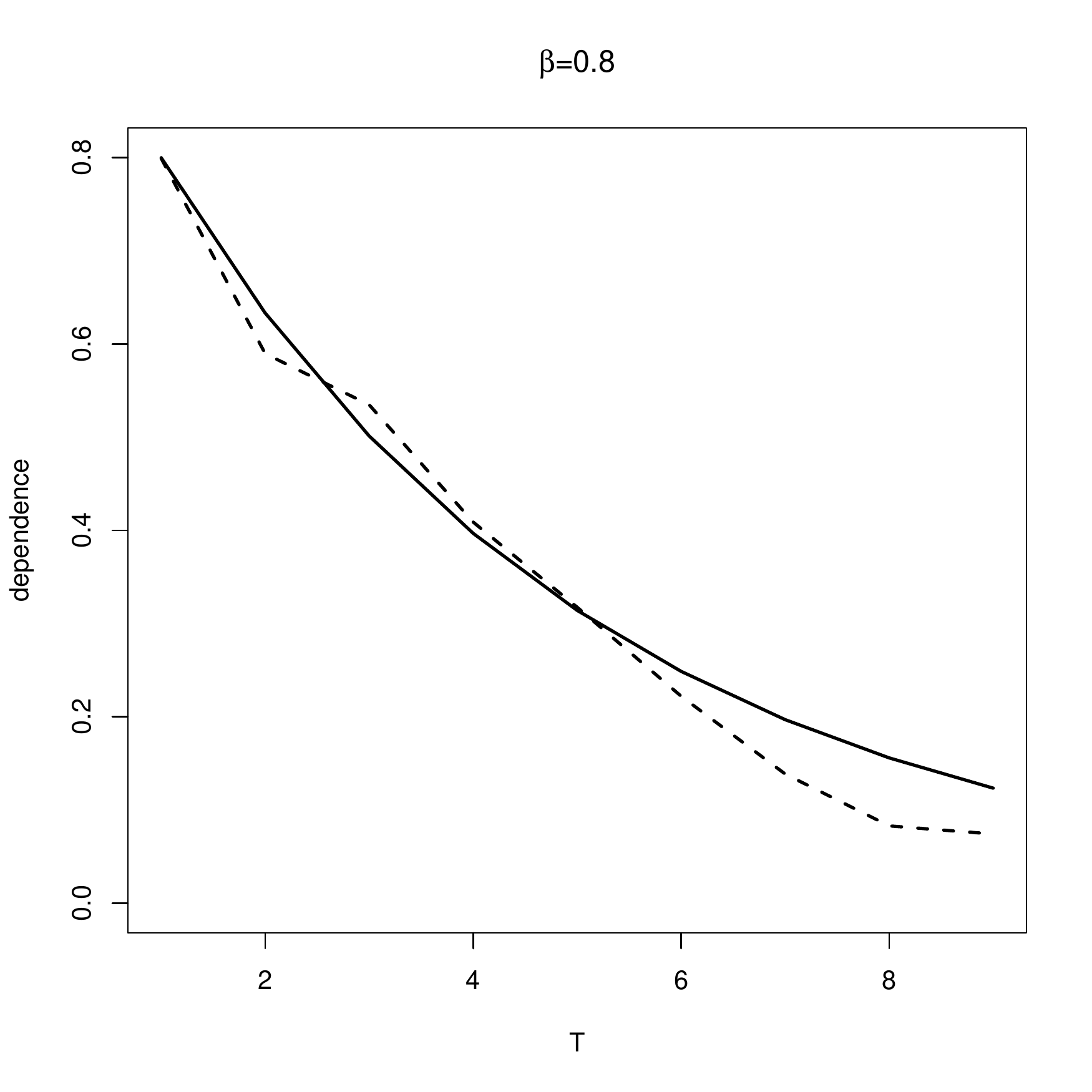}
  \caption{Plot of $\alpha^T\beta$ using RCoDA (solid line) against $\beta_T$ (dashed line), estimated by PL for the  $T^{th}$ sublattice, for a $q=2$ model over 256$\times$256 lattice, and at $\beta=0.5, 0.6, 0.7, 0.8$ from left to right.}
  \label{fig:decay}
 \end{figure}

To further investigate the appropriateness of using the decay rate of $\alpha^T\beta, 0<\alpha<1$ over the $T$ splits of the random field,  we separately estimated the value of $\beta_T$ for each $T^{th}$ sublattice using the full Potts model, PL was used to obtain the estimate for $\beta_T$. Figure \ref{fig:decay} shows the averaged estimate of $\beta_T$ and $\alpha^T\beta$ over 200 data sets simulated at $\beta=0.5, 0.6, 0.7, 0.8$. The model used here was a $q=2$ state Potts model over a 256$\times$256 lattice for different values of $\beta$. For very small lattice sizes or very weak dependences, the PL estimate is not reliable, possibly due to excessive boundary influence. Therefore, we show the decay for $T$ up to 8, corresponding to the smallest estimated lattice size of 16x16. Figure \ref{fig:decay} shows the curve $\alpha^T\beta$ and $\beta_T$ for
$T=0, \ldots, 8$, with the true $\beta=0.5, 0.6, 0.7, 0.8$. The graphs show a good match between the estimated $\beta_T$ and $\alpha^T\beta$, suggesting such a decay structure is appropriate.

\begin{figure}[ht]
 \centering
  \includegraphics[width=7cm,height=6cm]{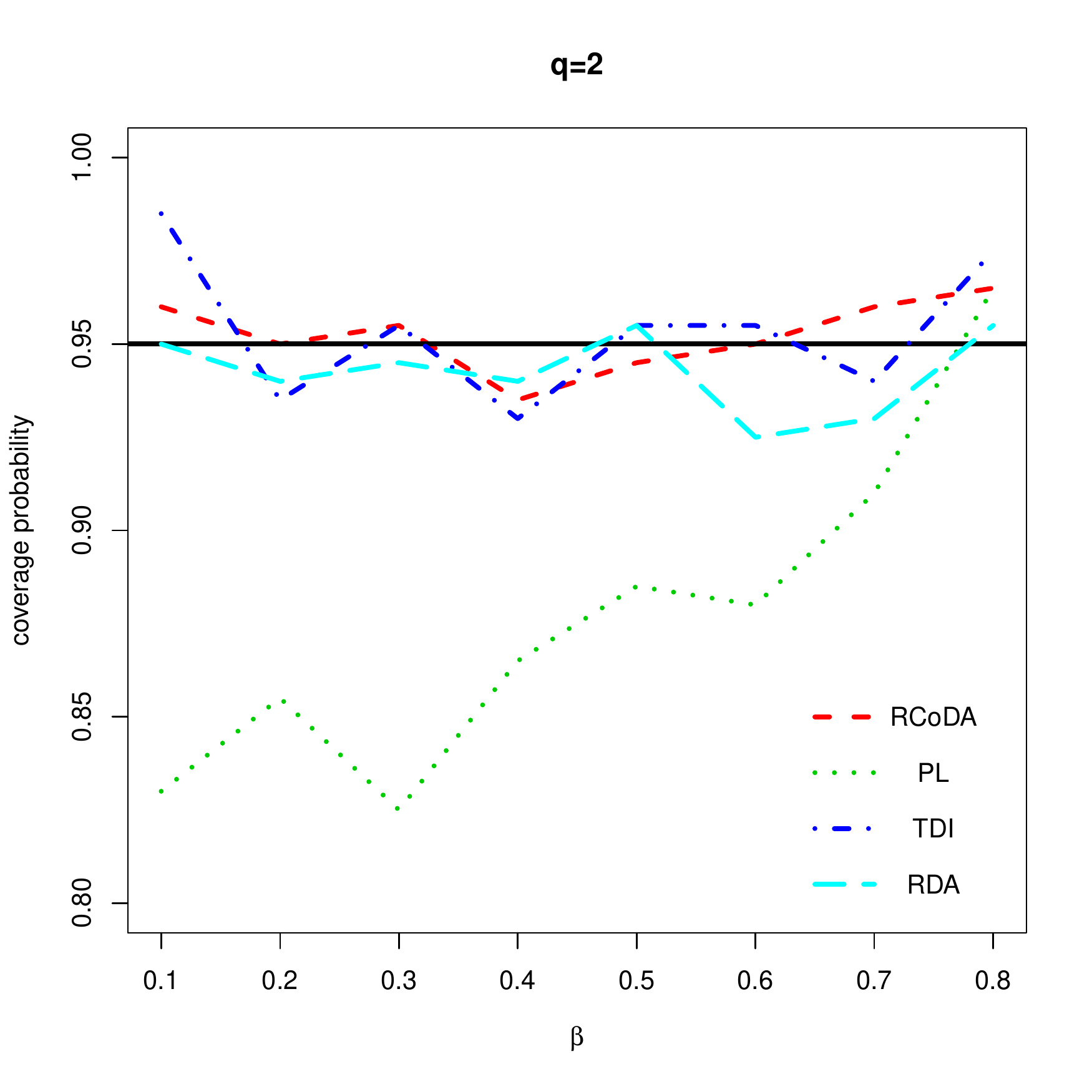}
    \includegraphics[width=7cm,height=6cm]{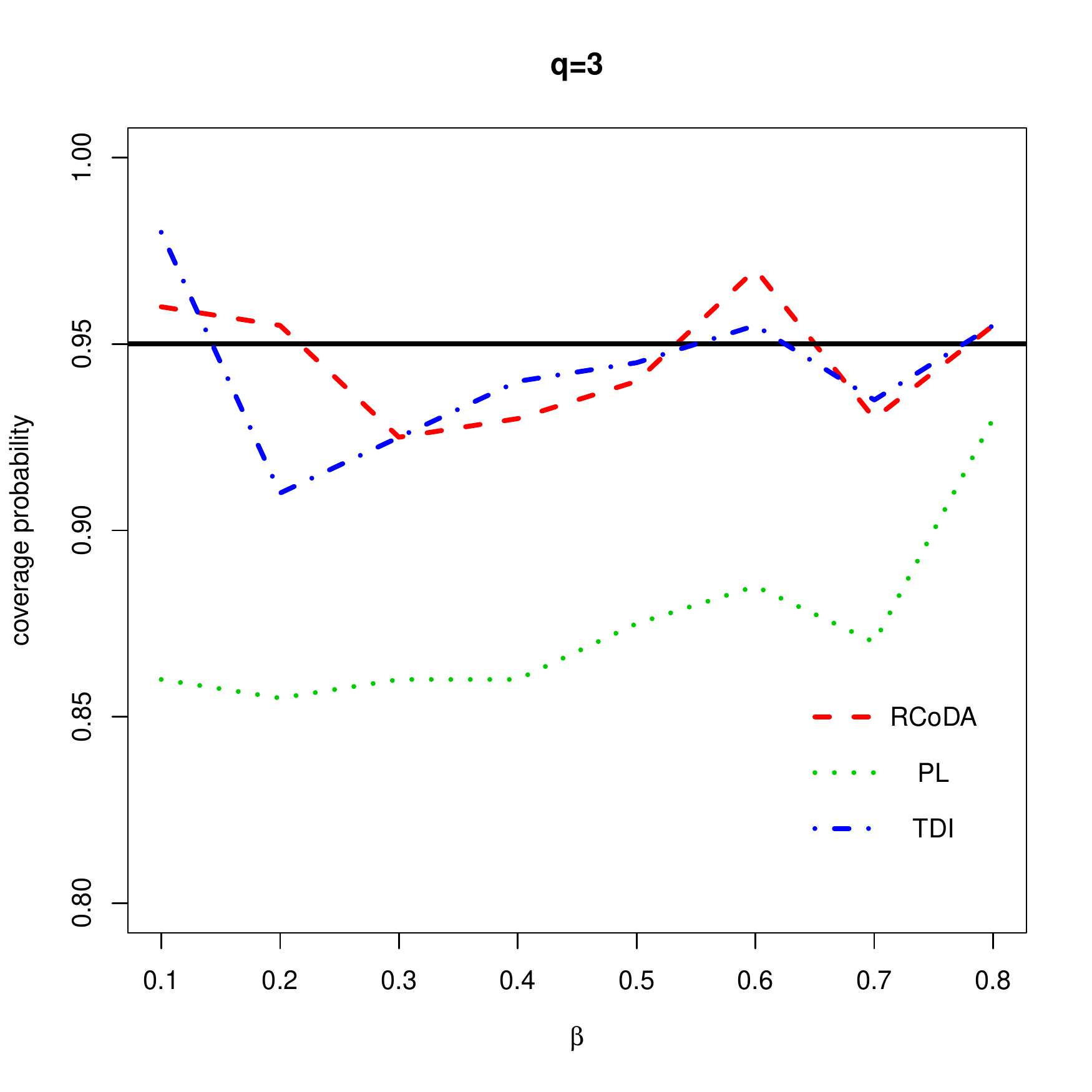}
  \caption{95\% empirical coverage probabilities for the 32 $\times$ 32 lattice with a first order neighbourhood, $q=2$ (left) and $q=3$ (right).}
  \label{fig:covfirst}
 \end{figure}

Figure \ref{fig:covfirst} shows the 95\% empirical coverage probabilities, estimated over varying values of $\beta$ and for $q=2$ and $q=3$ on a 32$\times$32 lattice.  For a given value of $\beta$, we simulated 200 datasets based on $\beta$. For each dataset, a 95\% posterior credibility interval of $\beta$ is recorded and the proportion of intervals containing the initial value of $\beta$ was recorded.
 It can be seen that the coverage probabilities of RCoDA, TDI and RDA are all close to the nominal level, suggesting that these methods yield valid inferences, see \shortciteN{monahanboos92}.
For TDI, this is expected, since the likelihood is exact. However, the coverage of PL is noticeably smaller than the nominal level, particularly at the weaker dependences. The phenomenon also corresponds to a generally narrower posterior variance estimate from our
simulation
results (not shown here). This is unsurprising since the pseudo-likelihood is a special case of composite likelihoods,  and direct computation using MCMC can result in posterior variances that are too small, see  \shortciteN{varin2011overview} and \shortciteN{pauli11} for discussions.

\subsection{Second order neighbourhood}
\begin{figure}[ht]
 \centering
\begin{subfigure}{0.22\textwidth}
               \includegraphics[width=\textwidth,height=0.15 \textheight]{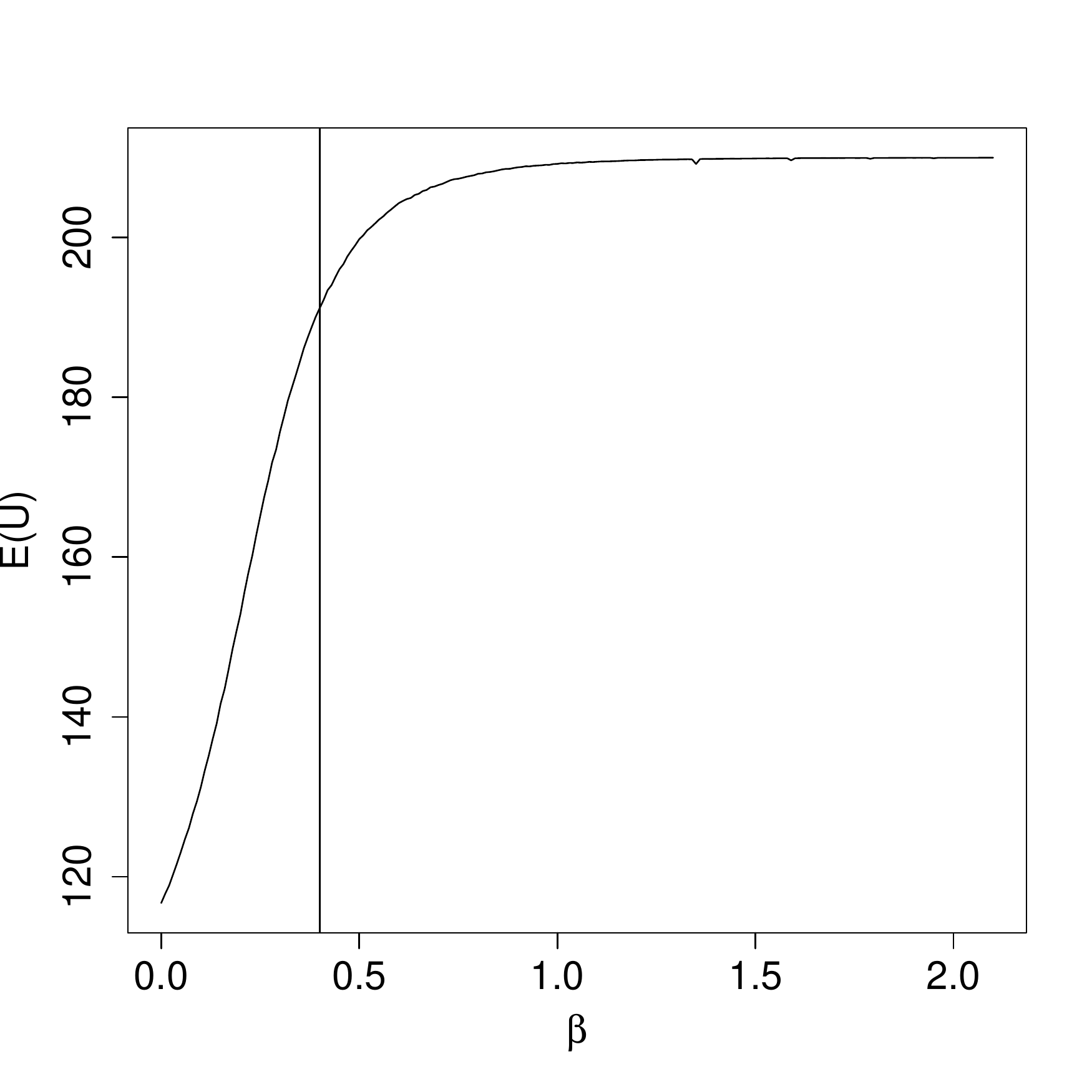}
               \caption{$8\times 8$ MRF}
\end{subfigure}
\begin{subfigure}{0.22\textwidth}
               \includegraphics[width=\textwidth,height=0.15 \textheight]{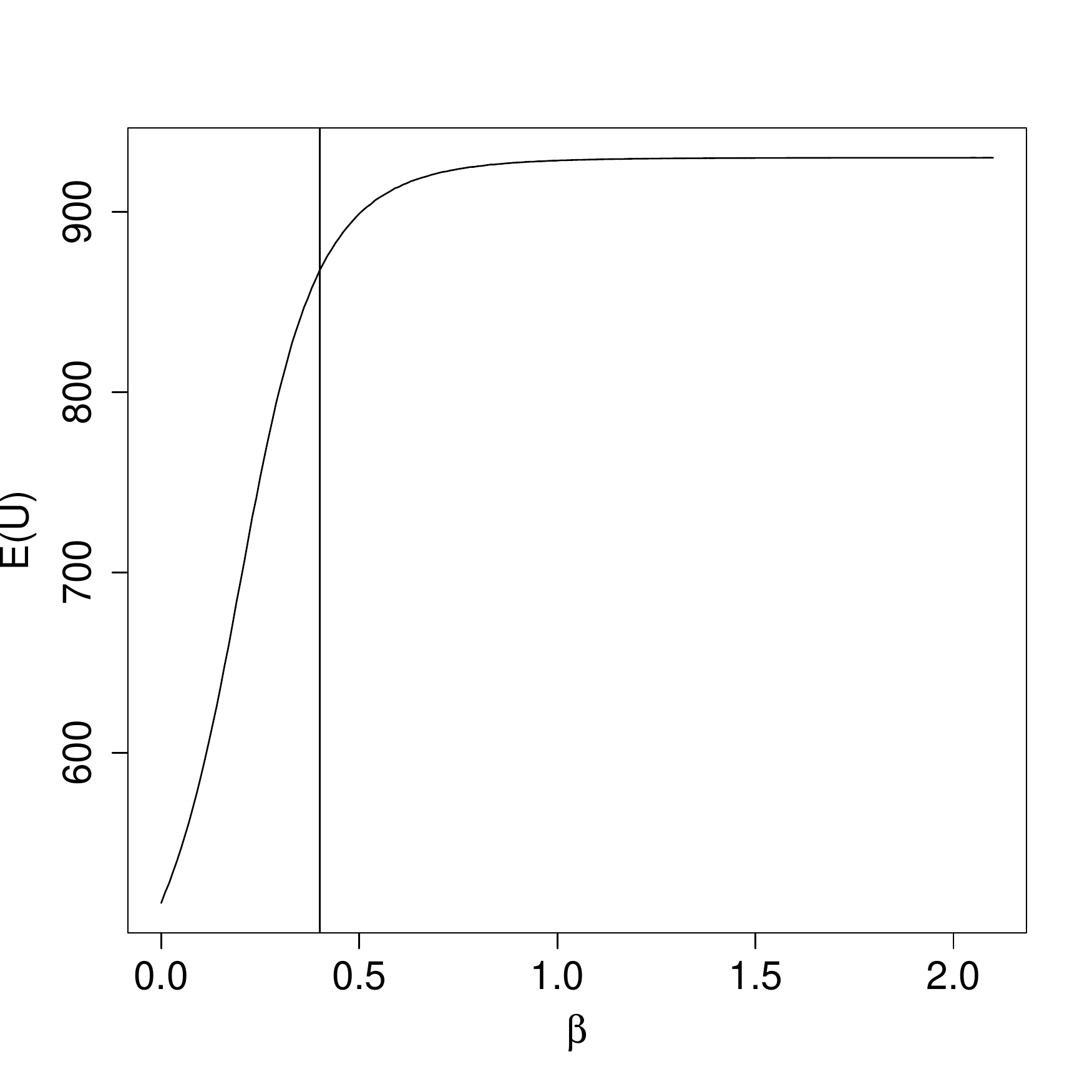}
               \caption{$16\times 16$ MRF}
\end{subfigure}
\begin{subfigure}{0.22\textwidth}
               \includegraphics[width=\textwidth,height=0.15 \textheight]{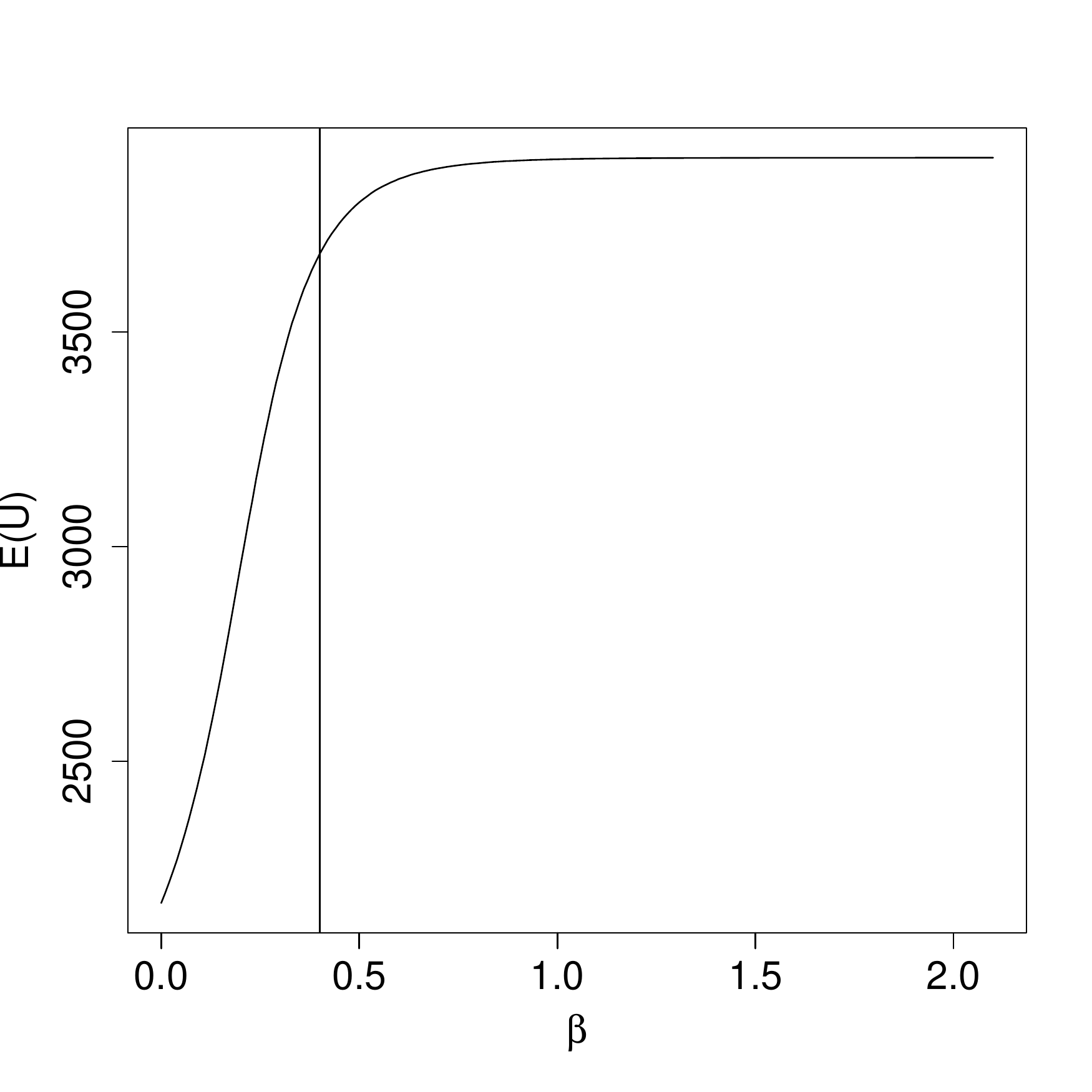}
               \caption{$32\times 32$ MRF}
\end{subfigure}
\begin{subfigure}{0.22\textwidth}
                \includegraphics[width=\textwidth,height=0.15 \textheight]{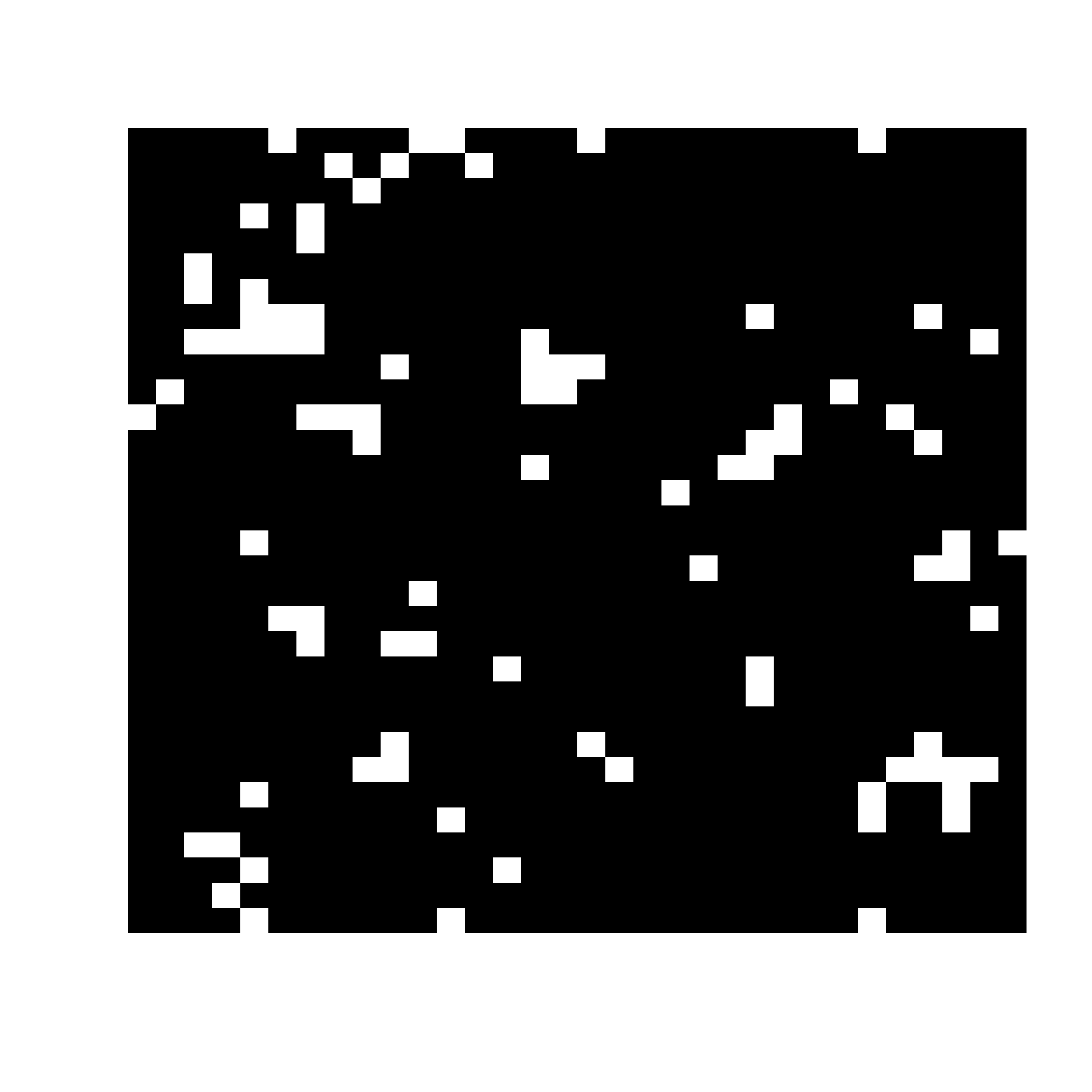}
                \caption{Ising model}
 \end{subfigure}
\caption{Estimates of $E(U(\bf z)|\beta)$for Ising model over different lattice size: (a) 8$\times$8 (b) 16$\times$16 and (c) 32$\times$32.
Vertical line correspond to  $\beta=0.4$. (d) shows simulation of one realization of the Ising model at $\beta=0.4$.}
 \label{petereusecond}
\end{figure}

For the second order neighbourhood study, we again considered the $q=2$ and $q=3$ state Potts model over  32$\times$32, 128$\times$128 and 256$\times$256 lattices. The RDA method was omitted here.
In order to determine the critical value for $\beta$, we monitored the changes in the value of $E(U(\bf z)|\beta)$, where $U({\bf z})=\sum_{i\sim j}I( z_i=z_j)$.  $U(\bf z)$ is the total number of pairs in $\bf z$. Figures \ref{petereusecond}(a)-(c) presents the changes in $E(U(\bf z)|\beta)$ as $\beta$ changes, for a number of different sizes of lattices. The estimated value of $E(U(\bf z)|\beta)$ was obtained by Monte Carlo method similar to that used for TDI.
It can be seen that the estimates stabilise around 0.4. Figure \ref{petereusecond}(d) presents one realization of the Ising model at $\beta=0.4$, where the figure
begins to be dominated by one colour, which is a sign of phase transition. Therefore, we restrict our study to $\beta < 0.4$.
See also \shortciteN{green2002hidden}, \shortciteN{gelman1998simulating} and \shortciteN{moores2015scalable} who discusses the uses of  $E(U(\bf z)|\beta)$ in inference.

Table \ref{infer-second} shows the root mean squared errors of the $\beta$ estimation for  $q=2$ and 3
over the varying lattice sizes, using RCoDA-C, RCoDA-M, PL and TDI. The results were computed over 200 simulated data sets at $\beta=0.1, 0.2$ and 0.3. The results suggest no significant difference in performance over the values of $q$. For the larger lattices, RCoDA-C, PL and TDI all performed similarly in terms of root mean squared errors. RCoDA-M, which assumes marginal independence, was worse overall compared to RCoDA-C, which uses a partial pseudo-likelihood. For the 32$\times$32 lattice, RCoDA methods performed worst, this suggests that it is not suitable to use decomposition in second order neighbourhoods when lattice sizes are too small, since the method of splitting requires that we should have at least several iterations. So when the lattice size is too small, the relative bias will be larger.

\begin{figure}[ht]
 \centering
  \includegraphics[width=7cm,height=6cm]{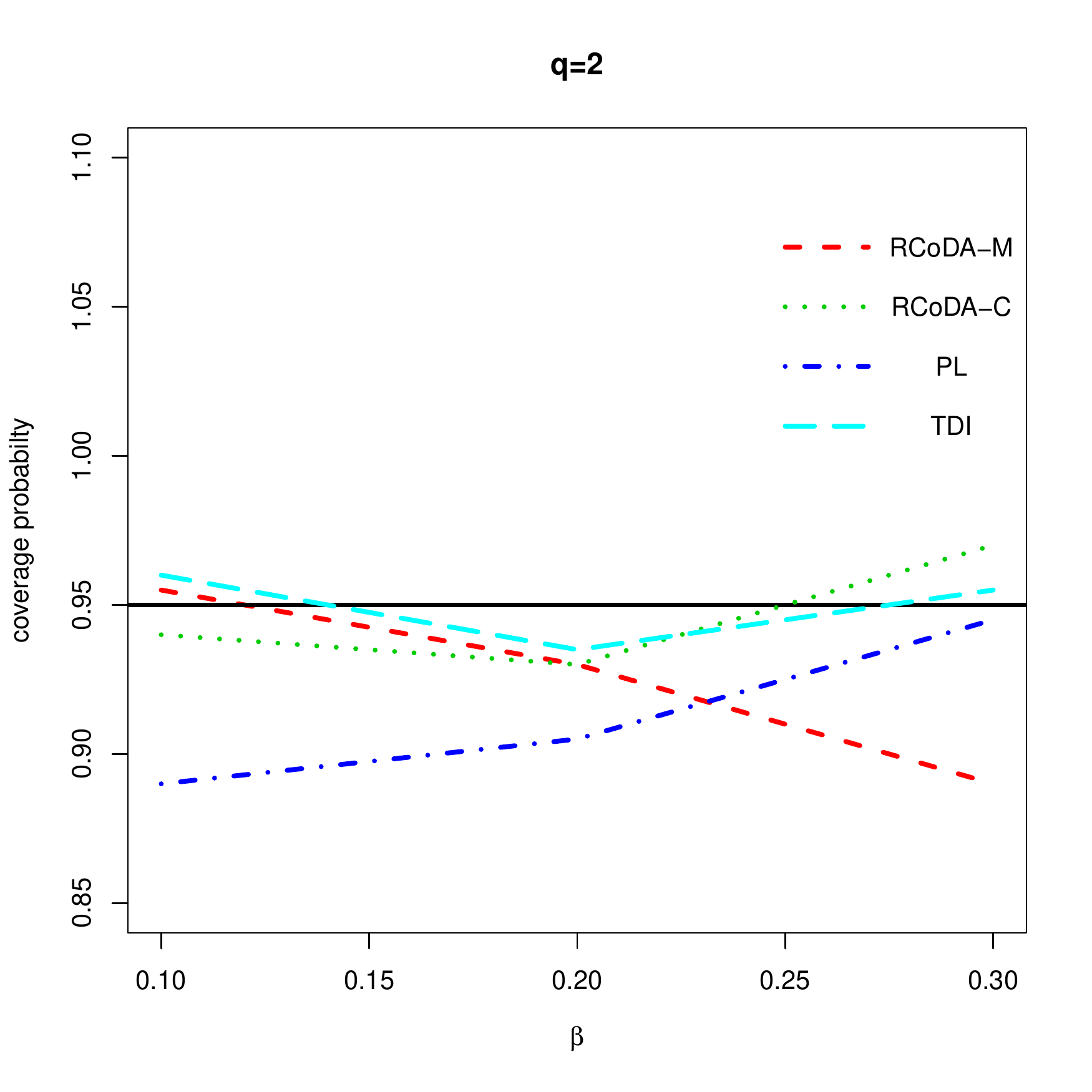}
  \includegraphics[width=7cm,height=6cm]{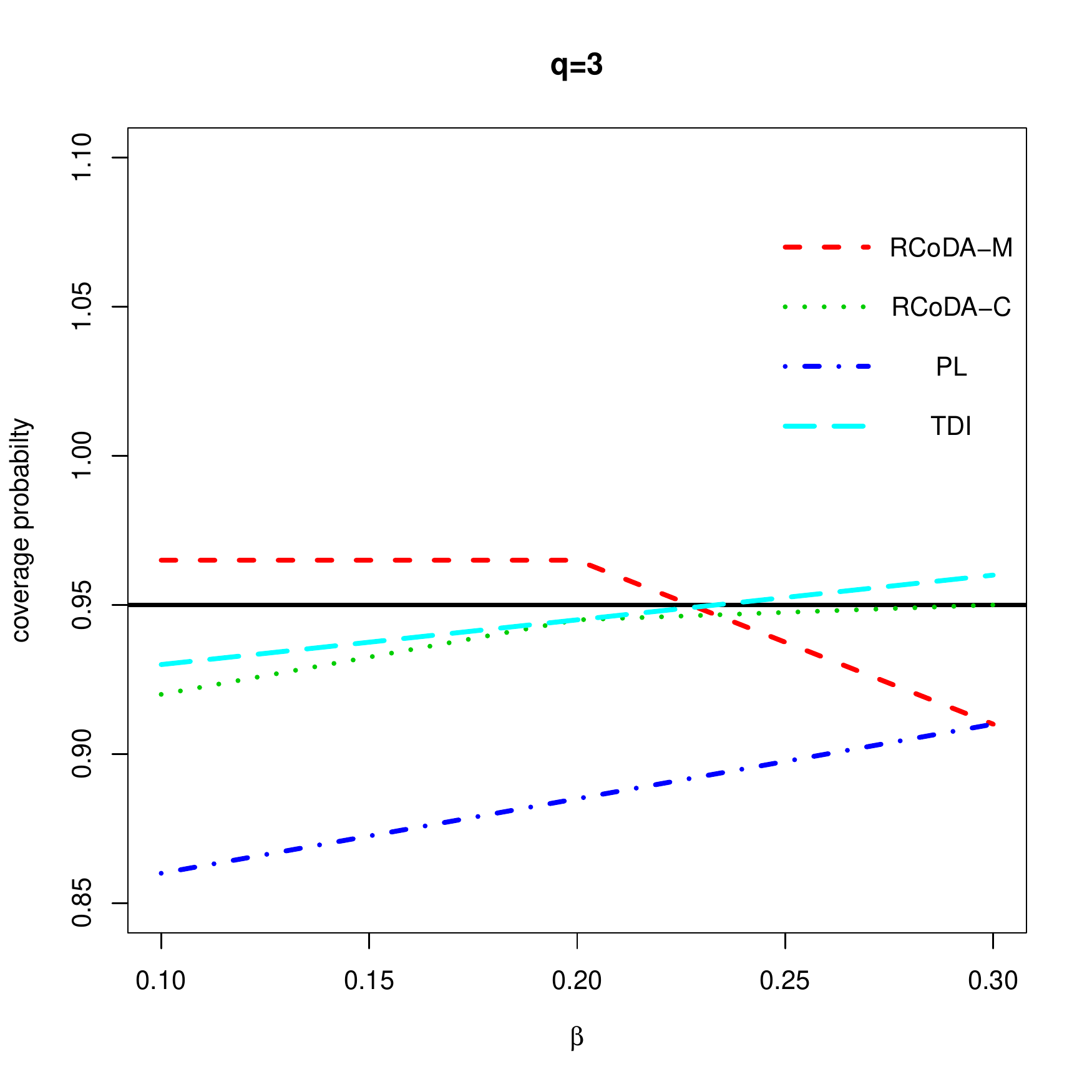}
  \caption{95\% empirical coverage probabilities for the 32 $\times$ 32 lattice with a second order neighbourhood, $q=2$ (left) and $q=3$ (right).}
  \label{fig:covsecond}
 \end{figure}

Figure \ref{fig:covsecond} shows the empirical coverage probabilities computed under similar conditions to those for first order neighbourhood simulations. Again, we see that the PL methods do not achieve good coverage, where as both RCoDA and TDI achieve good coverage, with RCoDA-C performing fairly consistently better.

\begin{table}[ht] \scriptsize
 \centering
   \begin{tabular}{|c|c|c|c|c|c|c|c|c|c|c|c|c|c|}
    \hline
    \hline
     && \multicolumn{4}{ c| }{$32^2$}  &\multicolumn{4}{ c| }{$128^2$} & \multicolumn{4}{ c| }{$256^2$}\\
     \cline{3-14}
          $\beta$ &$q$ & \tiny{RCoDA-M} & \tiny{RCoDA-C} & \tiny{PL }& \tiny{TDI} & \tiny{RCoDA-M} & \tiny{RCoDA-C} & \tiny{PL} & \tiny{TDI} & \tiny{RCoDA-M} & \tiny{RCoDA-C} & \tiny{PL}& \tiny{TDI}\\
    \hline
    \multirow{2}{*}{\tiny 0.1} &2 &0.029     & 0.029    & 0.026  & 0.025  & 0.008     &0.008      & 0.003   &0.006 &0.005  & 0.004    &0.003 &0.003\\
                    &3 &0.031     & 0.033    & 0.029  & 0.027  &0.004&0.004&0.003     & 0.007 &0.004     & 0.004    & 0.003 &0.003\\
    \hline

     \multirow{2}{*}{\tiny 0.2} &2 & 0.036    & 0.031    &0.025   & 0.024 & 0.015     & 0.007     & 0.003   &0.006 &0.013  & 0.004   &0.003  &0.003 \\
                    &3 & 0.031    & 0.029    &0.026   & 0.025 &0.009&0.004&0.003 & 0.006 & 0.009    & 0.004    &0.003 &0.003  \\
    \hline
    \multirow{2}{*}{\tiny 0.3} &2 & 0.038   & 0.027    & 0.021  & 0.018 & 0.032     & 0.007     & 0.002    &0.004 &0.031   & 0.004  &0.002 &0.002\\
                    &3 & 0.031   & 0.025    & 0.020  & 0.019 &0.023&0.004&0.003 &0.004 & 0.023   & 0.004    & 0.003  & 0.002\\
     \hline
   \end{tabular}
  \caption{Root mean squared error of $\beta$ for a second order neighbourhood dependence.  Based on 200 simulated data sets for each 32$\times$32, 128$\times$128 and 256$\times$256  lattices.  $q=2$ and $q=3$.}
 \label{infer-second}
\end{table}

\section{Real data application}
\label{sec:application}

\begin{figure} [ht]
 \centering
 \includegraphics[width=0.6\linewidth,height=0.25\textheight]{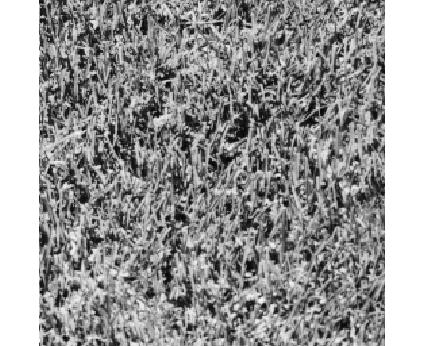}
 \caption{Grass image.}
 \label{fig:grass}
\end{figure}

We now apply our algorithm to an image of grass, which has been widely studied in texture modelling. The images are available online,  at  \url{http://sipi.usc.edu/database/database.php?volume=textures}. The image was originally studied in \shortciteN{brodatz1966textures}. Without loss generality, we take the first 256 rows and 256 columns as our data of interest, see Figure \ref{fig:grass}.

We use a two-component Gaussian mixture model to model the grass data.
The posterior distribution is given in Equation \ref{eq:mix}, with $\pi(y_i|\theta, z_i) $ given by the component Normal distribution according to $z_i$, with parameters $\mu_j$ and $\sigma_j, j=1,2$ indicating the component mean and variance. The distribution of ${\bf z}$ is the Ising model as given in Equation \ref{eq:potts}. We set prior distributions for $\mu_j \sim N(0.5, 100^2)$ and $ \sigma _j ^2 \sim IG(0.001,0.001), j=1,2$. The prior of $\beta$ is set to to Uniform distribution between 0 to 4.
 A Metropolis-Hastings algorithm was used in the MCMC.   6000 MCMC iterations was implemented, while the first 2000 iterations were thrown away as burn-in.  We fitted the Ising model
with first order and second order neighbourhood structure respectively. The results are presented in Table \ref{tab:grass inference}.

\begin{table} [ht]
 \centering
 \begin{tabular}{|c|c|c|c|c|c|}
  \hline
  \hline
     & $\mu_1$ & $\mu_2$ & $\sigma_1$ & $\sigma_2$ &  $\beta$ \\
     \hline
   \multirow{2} {*} {PL-(F)}& 0.251 & 0.609&  0.013& 0.019&  1.364\\
                          & (0.0022)& (0.0015)& (0.00026)&(0.00027)&(0.017)\\ \hline
   \multirow{2} {*}{RCoDA-(F)} & {0.265} & 0.620&  0.014& 0.018&  {1.280}\\
                          & (0.0028)& (0.0017)& (0.00037)&(0.00029)&(0.022)\\ \hline
   \multirow{2} {*}{TDI-(F)} & 0.302 & 0.650&  0.017& 0.013&  0.841\\
                          & (0.0015)& (0.0010)& (0.00022)&(0.00014)&(0.0033)\\ \hline
   \multirow{2} {*}{PL-(S)}&0.236&0.599 & 0.011 &0.021& 0.600 \\
                         &(0.0022) &(0.0015)&(0.00027)&(0.00029) &(0.0066) \\ \hline
   \multirow{2} {*}{RCoDA-C-(S)} & 0.252 & 0.611&0.013 &0.019 & 0.567 \\
                            &(0.0025)&(0.0016)&(0.00029) &(0.00030) & (0.0080) \\
                            \hline
    \multirow{2} {*}{TDI-(S)} & 0.303 & 0.649&0.017 &0.013 & 0.373 \\
                            &(0.0015)&(0.0010)&(0.00021) &(0.00016) & (0.0013) \\
   \hline
   \hline
 \end{tabular}
\caption{Posterior mean and standard deviation (in brackets) of grass data using PL, RCoDA and TDI respectively. (F) denotes first order neighbourhood structure. (S) denotes  second order neighbourhood structure. }
\label{tab:grass inference}
\end{table}

Table \ref{tab:grass inference} presents the posterior mean and standard deviation of the two-component Gaussian spatial mixture model using TDI and RCoDA (only RCoDA-C was implemented for the second order neighbourhood) and PL. For both neighbourhood structures, the estimates for $\beta$ were considerably different between the three methods, although the component mixture parameters were fairly similar. In both cases, PL gave the largest estimate for $\beta$, followed by RCoDA and TDI always produced smaller estimates for $\beta$. Since for simulated data, where we know that the data comes from the Potts model, the results produced by the three methods were very similar, this suggests that the grass image may not closely follow a Potts model. However, since we do not know the truth, the effect of the three different methods becomes difficult to evaluate.

In order to assess the estimation from the three different approaches, we consider the use of posterior predictive distributions. For each posterior sample, we can simulate an image dataset, consequently, for each pixel, we can compare the observed value of that pixel with the posterior predictive distribution for that pixel. Table \ref{tab:grass posterior} shows the percentage of observed pixels which fall within a 95\%, 90\% and 80\% of the posterior predictive distributions. We can see that here the three methods are quite similar, RCoDA having the higher proportions in most cases, indicative of a slightly better performance.

This example illustrates that for real datasets, the effect of possible model misspecification has different implications depending on the computational methods used. While a posterior predictive check appears to suggest all the methods are performing similarly, the posterior parameter estimates are quite different. This illustrates the importance of model checking and validation in this type of applications.

\begin{table}
 \centering
 \begin{tabular}{|c|c|c|c|c|c|c|}
  \hline
  \hline
  & PL-(F)& RCoDA-(F)&TDI-(F)& PL-(S)& RCoDA-(S)&TDI-(S) \\
  \hline
  95\% & 99.35& 99.54&98.97 &99.41& 99.35&99.03  \\
  \hline
   90\% &96.77 &97.24 &96.11 & 97.16&97.08 &96.18 \\ \hline
    80\% &87.04 & 88.11&88.88 & 87.67 &88.10 &89.03 \\ \hline  \hline
 \end{tabular}
\caption{Percentages of observed pixels which fall within the 95\%, 90\% and 80\% of the posterior predictive distributions.}
\label{tab:grass posterior}
\end{table}

\section{Discussions}
\label{sec:conclusion}

In this article we have proposed a new method of estimating the $q$-state Potts model without having to compute the usually intractable normalising constant. Our method recursively partitions a regular lattice into a conditionally independent sublattice and approximates the other by another Potts model with a weaker dependence. By doing so, the method effectively avoids the computation of the troublesome normalising constant. We presented the method in terms of first and second order neighbourhood structure on a 2D lattice. More complex lattices and dependence structures may be possible but would be much more difficult to work with.  The method was demonstrated for $q=2$ and $q=3$ in this article, but can be applied to any $q$.

The proposed method is computationally efficient, the computational complexity is of the same order of magnitude as that of PL. We have shown through our simulation studies that RCoDA obtains the correct empirical coverage probabilities, whereas PL does not always do so. We have shown that for the first order neighbourhood, the estimation in terms of root mean squared error is competitive with
several existing methods for different values of $q$ and lattice sizes. For the second order neighbourhood structure, RCoDA produces better results when the size of the lattice is large.

 The advantage of RCoDA over other methods such as RDA and TDI, is its scalability in $q$ and lattice size. Extensions of the proposed RCoDA methods may be to consider variants of Potts models, such as the Potts model with an external field.

\newpage
\bibliographystyle{chicago}
\bibliography{normalc}

\begin{thebibliography}{}

\bibitem[\protect\citeauthoryear{Aizenman, Chayes, Chayes, and Newman}{Aizenman
  et~al.}{1988}]{aizenman1988discontinuity}
Aizenman, M., J.~Chayes, L.~Chayes, and C.~Newman (1988).
\newblock Discontinuity of the magnetization in one-dimensional {$1/ |x- y|^2$
  } {I}sing and {P}otts models.
\newblock {\em Journal of Statistical Physics\/}~{\em 50\/}(1-2), 1--40.

\bibitem[\protect\citeauthoryear{Barkema and de~Boer}{Barkema and
  de~Boer}{1991}]{barkema1991numerical}
Barkema, G. and J.~de~Boer (1991).
\newblock Numerical study of phase transitions in {P}otts models.
\newblock {\em Physical Review A\/}~{\em 44\/}(12), 8000--8005.

\bibitem[\protect\citeauthoryear{Bartolucci and Besag}{Bartolucci and
  Besag}{2002}]{bartolucci2002recursive}
Bartolucci, F. and J.~Besag (2002).
\newblock A recursive algorithm for {M}arkov random fields.
\newblock {\em Biometrika\/}~{\em 89\/}(3), 724--730.

\bibitem[\protect\citeauthoryear{Besag}{Besag}{1974}]{besag1974spatial}
Besag, J. (1974).
\newblock Spatial interaction and the statistical analysis of lattice systems.
\newblock {\em Journal of the Royal Statistical Society. Series B
  (Methodological)\/}~{\em 36\/}(2), 192--236.

\bibitem[\protect\citeauthoryear{Brodatz}{Brodatz}{1966}]{brodatz1966textures}
Brodatz, P. (1966).
\newblock {\em Textures: a photographic album for artists and designers},
  Volume~66.
\newblock New York: Dover.

\bibitem[\protect\citeauthoryear{Celeux, Forbes, and Peyrard}{Celeux
  et~al.}{2003}]{celeux2003procedures}
Celeux, G., F.~Forbes, and N.~Peyrard (2003).
\newblock {EM} procedures using mean field-like approximations for {M}arkov
  model-based image segmentation.
\newblock {\em Pattern recognition\/}~{\em 36\/}(1), 131--144.

\bibitem[\protect\citeauthoryear{Cressie and Davidson}{Cressie and
  Davidson}{1998}]{cressie1998image}
Cressie, N. and J.~L. Davidson (1998).
\newblock Image analysis with partially ordered {M}arkov models.
\newblock {\em Computational statistics \& data analysis\/}~{\em 29\/}(1),
  1--26.

\bibitem[\protect\citeauthoryear{Cressie and Cassie}{Cressie and
  Cassie}{1993}]{cressie1993statistics}
Cressie, N.~A. and N.~A. Cassie (1993).
\newblock {\em Statistics for spatial data}, Volume 900.
\newblock New York: Wiley.

\bibitem[\protect\citeauthoryear{Everitt}{Everitt}{2012}]{everitt2012bayesian}
Everitt, R.~G. (2012).
\newblock {B}ayesian parameter estimation for latent {M}arkov random fields and
  social networks.
\newblock {\em Journal of Computational and graphical Statistics\/}~{\em
  21\/}(4), 940--960.

\bibitem[\protect\citeauthoryear{Feng}{Feng}{2008}]{feng2008bayesian}
Feng, D. (2008).
\newblock {\em Bayesian hidden {M}arkov normal mixture models with application
  to {MRI} tissue classification}.
\newblock Ph.\ D. thesis, University of Iowa.

\bibitem[\protect\citeauthoryear{Feng, Tierney, and Magnotta}{Feng
  et~al.}{2012}]{feng2012mri}
Feng, D., L.~Tierney, and V.~Magnotta (2012).
\newblock {MRI} tissue classification using high-resolution {B}ayesian hidden
  {M}arkov normal mixture models.
\newblock {\em Journal of the American Statistical Association\/}~{\em
  107\/}(497), 102--119.

\bibitem[\protect\citeauthoryear{Friel, Pettitt, Reeves, and Wit}{Friel
  et~al.}{2009}]{friel2009bayesian}
Friel, N., A.~Pettitt, R.~Reeves, and E.~Wit (2009).
\newblock {B}ayesian inference in hidden markov random fields for binary data
  defined on large lattices.
\newblock {\em Journal of Computational and Graphical Statistics\/}~{\em
  18\/}(2), 243--261.

\bibitem[\protect\citeauthoryear{Gelman and Meng}{Gelman and
  Meng}{1998}]{gelman1998simulating}
Gelman, A. and X.-L. Meng (1998).
\newblock Simulating normalizing constants: from importance sampling to bridge
  sampling to path sampling.
\newblock {\em Statistical science\/}~{\em 13\/}(2), 163--185.

\bibitem[\protect\citeauthoryear{Geyer and Thompson}{Geyer and
  Thompson}{1992}]{geyer1992constrained}
Geyer, C.~J. and E.~A. Thompson (1992).
\newblock Constrained {M}onte {C}arlo maximum likelihood for dependent data
  (with discussion).
\newblock {\em Journal of the Royal Statistical Society. Series B.
  Methodological\/}~{\em 54\/}(3), 657--699.

\bibitem[\protect\citeauthoryear{Green and Richardson}{Green and
  Richardson}{2002}]{green2002hidden}
Green, P.~J. and S.~Richardson (2002).
\newblock Hidden {M}arkov models and disease mapping.
\newblock {\em Journal of the American statistical association\/}~{\em
  97\/}(460), 1055--1070.

\bibitem[\protect\citeauthoryear{Gu and Zhu}{Gu and Zhu}{2001}]{gu2001maximum}
Gu, M.~G. and H.-T. Zhu (2001).
\newblock Maximum likelihood estimation for spatial models by {M}arkov chain
  {M}onte {C}arlo stochastic approximation.
\newblock {\em Journal of the Royal Statistical Society: Series B (Statistical
  Methodology)\/}~{\em 63\/}(2), 339--355.

\bibitem[\protect\citeauthoryear{Hurn, Husby, and Rue}{Hurn
  et~al.}{2003}]{hurn2003tutorial}
Hurn, M.~A., O.~K. Husby, and H.~Rue (2003).
\newblock A tutorial on image analysis.
\newblock In J.~M{\o}ller (Ed.), {\em Spatial statistics and computational
  methods}, Volume 173 of {\em Lecture Notes in Statistics}, pp.\  87--141. New
  York: Springer.

\bibitem[\protect\citeauthoryear{Knorr-Held and Rue}{Knorr-Held and
  Rue}{2002}]{knorr2002block}
Knorr-Held, L. and H.~Rue (2002).
\newblock On block updating in {M}arkov random field models for disease
  mapping.
\newblock {\em Scandinavian Journal of Statistics\/}~{\em 29\/}(4), 597--614.

\bibitem[\protect\citeauthoryear{Kosterlitz}{Kosterlitz}{1974}]{kosterlitz1974critical}
Kosterlitz, J. (1974).
\newblock The critical properties of the two-dimensional xy model.
\newblock {\em Journal of Physics C: Solid State Physics\/}~{\em 7\/}(6),
  1046--1060.

\bibitem[\protect\citeauthoryear{Li and Singh}{Li and
  Singh}{2009}]{li2009markov}
Li, S.~Z. and S.~Singh (2009).
\newblock {\em {M}arkov random field modeling in image analysis}, Volume~3 of
  {\em Advances in Pattern Recognition}.
\newblock London: Springer.

\bibitem[\protect\citeauthoryear{Liang}{Liang}{2007}]{liang2007continuous}
Liang, F. (2007).
\newblock Continuous contour {M}onte {C}arlo for marginal density estimation
  with an application to a spatial statistical model.
\newblock {\em Journal of Computational and Graphical Statistics\/}~{\em
  16\/}(3), 608--632.

\bibitem[\protect\citeauthoryear{Liang}{Liang}{2010}]{liang2010double}
Liang, F. (2010).
\newblock A double {M}etropolis-{H}astings sampler for spatial models with
  intractable normalizing constants.
\newblock {\em Journal of Statistical Computation and Simulation\/}~{\em
  80\/}(9), 1007--1022.

\bibitem[\protect\citeauthoryear{Liang, Jin, Song, and Liu}{Liang
  et~al.}{2015}]{liangjsl15}
Liang, F., I.~H. Jin, Q.~Song, and J.~S. Liu (2015).
\newblock An adaptive exchange algorithm for sampling from distributions with
  intractable normalising constants.
\newblock {\em Journal of the American Statistical Association\/}~{\em In
  Press}.

\bibitem[\protect\citeauthoryear{Lindsay}{Lindsay}{1988}]{lindsay1988composite}
Lindsay, B.~G. (1988).
\newblock Composite likelihood methods.
\newblock {\em Contemporary Mathematics\/}~{\em 80\/}(1), 221--39.

\bibitem[\protect\citeauthoryear{Luijten and Bl{\"o}te}{Luijten and
  Bl{\"o}te}{1995}]{luijten1995monte}
Luijten, E. and H.~W. Bl{\"o}te (1995).
\newblock Monte carlo method for spin models with long-range interactions.
\newblock {\em International Journal of Modern Physics C\/}~{\em 6\/}(03),
  359--370.

\bibitem[\protect\citeauthoryear{Lyne, Girolami, Atchad{'}e, Strathmann, and
  Simpson}{Lyne et~al.}{2015}]{lynegass15}
Lyne, A.-M., M.~Girolami, Y.~Atchad{'}e, H.~Strathmann, and D.~Simpson (2015).
\newblock On {R}ussian roulette estimates for {B}ayesian inference with
  doubly-intractable likelihoods.
\newblock {\em Statistical science\/}~{\em 30\/}(4), 443--467.

\bibitem[\protect\citeauthoryear{M{\o}ller, Pettitt, Reeves, and
  Berthelsen}{M{\o}ller et~al.}{2006}]{moller2006efficient}
M{\o}ller, J., A.~N. Pettitt, R.~Reeves, and K.~K. Berthelsen (2006).
\newblock An efficient markov chain monte carlo method for distributions with
  intractable normalising constants.
\newblock {\em Biometrika\/}~{\em 93\/}(2), 451--458.

\bibitem[\protect\citeauthoryear{Monahan and Boos}{Monahan and
  Boos}{1992}]{monahanboos92}
Monahan, J.~F. and D.~D. Boos (1992).
\newblock Proper likelihoods for bayesian analysis.
\newblock {\em Biometrika\/}~{\em 79\/}(2), 271--278.

\bibitem[\protect\citeauthoryear{Moores, Pettitt, and Mengersen}{Moores
  et~al.}{2015}]{moores2015scalable}
Moores, M.~T., A.~N. Pettitt, and K.~Mengersen (2015).
\newblock Scalable {B}ayesian inference for the inverse temperature of a hidden
  {P}otts model.
\newblock {\em arXiv preprint arXiv:1503.08066\/}.

\bibitem[\protect\citeauthoryear{Murray}{Murray}{2007}]{murray07advancesin}
Murray, I. (2007).
\newblock {\em Advances in Markov chain Monte Carlo methods}.
\newblock Ph.\ D. thesis, University of Cambridge.

\bibitem[\protect\citeauthoryear{Murray, Ghahramani, and MacKay}{Murray
  et~al.}{2006}]{murray06}
Murray, I., Z.~Ghahramani, and D.~J.~C. MacKay (2006).
\newblock {MCMC} for doubly-intractable distributions.
\newblock In {\em Proceedings of the 22nd Annual Conference on Uncertainty in
  Artificial Intelligence (UAI-06)}, pp.\  359--366. AUAI Press.

\bibitem[\protect\citeauthoryear{Nott and Ryd{\'e}n}{Nott and
  Ryd{\'e}n}{1999}]{nott1999pairwise}
Nott, D.~J. and T.~Ryd{\'e}n (1999).
\newblock Pairwise likelihood methods for inference in image models.
\newblock {\em Biometrika\/}~{\em 86\/}(3), 661--676.

\bibitem[\protect\citeauthoryear{Pal and Pal}{Pal and
  Pal}{1993}]{pal1993review}
Pal, N.~R. and S.~K. Pal (1993).
\newblock A review on image segmentation techniques.
\newblock {\em Pattern recognition\/}~{\em 26\/}(9), 1277--1294.

\bibitem[\protect\citeauthoryear{Pauli, Racugno, and Ventura}{Pauli
  et~al.}{2011}]{pauli11}
Pauli, F., W.~Racugno, and L.~Ventura (2011).
\newblock {B}ayesian composite marginal likelihoods.
\newblock {\em Statistica Sinica\/}~{\em 21\/}(1), 149--164.

\bibitem[\protect\citeauthoryear{Potts}{Potts}{1952}]{potts1952some}
Potts, R.~B. (1952).
\newblock Some generalized order-disorder transformations.
\newblock In {\em Mathematical proceedings of the cambridge philosophical
  society}, Volume~48, pp.\  106--109. Cambridge Univ Press.

\bibitem[\protect\citeauthoryear{Propp and Wilson}{Propp and
  Wilson}{1998}]{proppwilson98}
Propp, J.~G. and D.~B. Wilson (1998).
\newblock Exact sampling with coupled {M}arkov chains and applications to
  statistical mechanics.
\newblock {\em Random structures and algorithms\/}~{\em 9}, 223--252.

\bibitem[\protect\citeauthoryear{Reeves and Pettitt}{Reeves and
  Pettitt}{2004}]{reeves2004efficient}
Reeves, R. and A.~N. Pettitt (2004).
\newblock Efficient recursions for general factorisable models.
\newblock {\em Biometrika\/}~{\em 91\/}(3), 751--757.

\bibitem[\protect\citeauthoryear{Van~Leemput, Maes, Vandermeulen, and
  Suetens}{Van~Leemput et~al.}{1999}]{van1999automated}
Van~Leemput, K., F.~Maes, D.~Vandermeulen, and P.~Suetens (1999).
\newblock Automated model-based tissue classification of {MR} images of the
  brain.
\newblock {\em Medical Imaging, IEEE Transactions on\/}~{\em 18\/}(10),
  897--908.

\bibitem[\protect\citeauthoryear{Varin, Reid, and Firth}{Varin
  et~al.}{2011}]{varin2011overview}
Varin, C., N.~M. Reid, and D.~Firth (2011).
\newblock An overview of composite likelihood methods.
\newblock {\em Statistica Sinica\/}~{\em 21\/}(1), 5--42.

\bibitem[\protect\citeauthoryear{Wilkinson}{Wilkinson}{2006}]{wilkinson2006parallel}
Wilkinson, D.~J. (2006).
\newblock Parallel {B}ayesian computation.
\newblock Volume 184, pp.\  477--513. Marcel Dekker Ag.

\bibitem[\protect\citeauthoryear{Winkler}{Winkler}{2003}]{winkler2003image}
Winkler, G. (2003).
\newblock {\em Image analysis, random fields and {M}arkov chain {M}onte {C}arlo
  methods: a mathematical introduction}, Volume~27 of {\em Stochastic Modelling
  and Applied Probability}.
\newblock Verlag Berlin Heidelberg: Springer.

\bibitem[\protect\citeauthoryear{Wu}{Wu}{1982}]{wu1982potts}
Wu, F.-Y. (1982).
\newblock The {P}otts model.
\newblock {\em Reviews of modern physics\/}~{\em 54\/}(1), 235--268.

\end{thebibliography}
\end{document}